%% file: KATRIN.tex
\documentclass[12pt]{article}
\pdfoutput=1 
\usepackage{amsmath,amssymb,amsthm}
\usepackage{bbm,latexsym}
\usepackage{graphicx}
\usepackage{rotating} 
\usepackage[numbers,sort&compress]{natbib}
\usepackage{authblk}
\usepackage{slashed} 
\usepackage{xcolor}
\newcommand{\bone}{\mathbbm{1}}

\newcommand{\hc}{\mathrm{H.c.}}
\newcommand{\mA}{m_\mathcal{A}}
\newcommand{\mB}{m_\mathcal{B}}
\newcommand{\col}{\textcolor{red!80!orange}}
\def\gs{\mathrel{
   \rlap{\raise 0.511ex \hbox{$>$}}{\lower 0.511ex \hbox{$\sim$}}}}
\def\ls{\mathrel{
   \rlap{\raise 0.511ex \hbox{$<$}}{\lower 0.511ex \hbox{$\sim$}}}}

\begin{document}
\title{
\LARGE \bf Direct Neutrino Mass Experiments and 
Exotic Charged Current Interactions}
\setcounter{footnote}{2}
\author{Patrick Otto Ludl\thanks{E-mail: patrick.ludl@mpi-hd.mpg.de}\quad}
\author{\quad Werner Rodejohann\thanks{E-mail: werner.rodejohann@mpi-hd.mpg.de}}
\affil{{\small Max-Planck-Institut f\"ur Kernphysik, Saupfercheckweg 1, D-69117 Heidelberg, Germany}}

\date{April 8, 2016}

\maketitle

\begin{abstract}
\noindent
We study the effect of exotic charged current interactions 
on the electron energy spectrum in tritium decay, focussing on 
the KATRIN experiment and a possible modified setup that has access to the full spectrum. Both 
sub-eV and keV neutrino masses are considered. We perform a fully 
relativistic calculation and take all possible new interactions into account, 
demonstrating the possible sizable distortions in the energy spectrum. 

\end{abstract}

\newpage
\input{introduction}

\input{relativistic}
\input{SM}
\input{SM_corr}
\input{NP}
\input{conclusions}

\paragraph{Acknowledgements:} P.O.L.\ thanks Hiren Patel and Xunjie Xu 
for helpful discussions and the MPIK Heidelberg for its hospitality and the
excellent working atmosphere. W.R.\ was supported by the DFG in the 
Heisenberg Programme with grant RO 2516/6-1. 
We thank Mart\'in Gonz\'alez-Alonso for valuable comments.

\input{appendix}

\input{bibliography}
\end{document}

%% file: introduction.tex
\section{\label{sec:intro}Introduction}

Studies of nuclear $\beta$-decay are a popular probe of 
physics beyond the 
Standard Model~\cite{Herczeg:2001vk,Severijns:2006dr,cirigliano,Vos:2015eba}. 
Interestingly, high precision studies of the (near endpoint) 
nuclear $\beta$-spectrum of tritium will 
soon be possible with the KATRIN experiment~\cite{Angrik:2005ep}. 
The main physics goal is to determine the value of the absolute 
neutrino mass if it lies above 0.35 eV, or to set strong limits going down 
to 0.2 eV, improving current direct constraints by an order of magnitude. 
The possible presence of exotic charged current 
interactions in direct neutrino mass experiments (see~\cite{Otten:2008zz,Drexlin:2013lha} 
for reviews) has often been analyzed~\cite{Shrock1,Shrock2,Shrock3,Stephenson:1998cx,Stephenson:2000mw,Ignatiev:2005nu,Bonn:2007su,Simkovic,Simkovic2,SejersenRiis:2011sj}. The outcome of such
investigations
is that with studies near the endpoint of the spectrum 
existing limits on exotic interactions cannot be much improved. Moreover, 
if the endpoint is left as free parameter in the analysis, 
the presence of new interactions will have only little effect on the 
neutrino mass determination~\cite{Bonn:2007su,SejersenRiis:2011sj}. 

However, it is possible to modify the KATRIN setup in order 
to access the whole spectrum. This has been proposed in refs.~\cite{tritium-sensitivity,tritium-sensitivity2}
as a possibility to look for sterile neutrinos 
with masses around a few keV, which are 
interesting values in terms of Warm Dark Matter~\cite{Adhikari:2016bei}. 
In general, with
KATRIN's $10^{11} $ tritium decays per second the experiment provides 
an excellent opportunity to look for
spectral distortions,
which are characteristic for new mass states but also for 
exotic interactions. 
Having this in mind, there have been already 
papers studying the presence of keV neutrinos in the
spectrum~\cite{tritium-sensitivity,deVega:2011xh,Rodejohann:2014eka}, 
which would leave a characteristic kink at $Q - m_S$ in the
electron energy spectrum 
($Q$ being the endpoint, $m_S$ the mass of the sterile neutrino).
The presence of an additional right-handed interaction, further 
modifying the shape of the spectrum, has also been studied~\cite{Barry}. 

The goal of the present paper is to perform a general analysis of the
electron energy spectrum
of tritium $\beta$-decay in the 
presence of exotic charged 
current interactions and to illustrate the possible spectral 
distortions that can be observed if the whole spectrum is accessible. 
In our
generic
relativistic calculation we find that regardless of the interaction the 
energy spectrum can be parameterized by six functions which depend only on the involved 
particle masses and coupling constants, and whose precise form is specified by the interaction.  
We use an effective operator approach to study all possible Lorentz-invariant 
charged current interactions~\cite{cirigliano,Cirigliano:2012ab} including right-handed sterile neutrinos. 
Both small neutrino masses of order 0.5 eV and large masses of order 
keV are considered. 
While the endpoint region does not display 
significant effects, the full spectrum can display sizable distortions
on the permille level, even for unobservably 
small neutrino masses. This allows in principle 
to improve the bounds on the effective operators and adds additional physics motivation to 
modifications of high activity neutrino mass experiments to study the full spectrum.\footnote{We note that 
the Project 8 experiment~\cite{Monreal:2009za} 
has in principle also access to the full 
spectrum and our results would apply in this case as well. 
The PTOLEMY project also discusses possible constraints on 
keV-scale neutrinos~\cite{Betts:2013uya}.}  \\

The paper is build up as follows: in section \ref{sec:rel} we study the most general 
electron energy spectrum of $\beta$-decay, working in a 
relativistic approach and keeping the underlying interaction 
unspecified. 
Using this general formalism, we revisit the 
Standard Model spectrum and Kurie-plot for tritium $\beta$-decay 
in section \ref{sec:SM}, where we also analyze the corrections to well-known 
textbook results originating from a proper relativistic treatment. In section 
\ref{sec:NP} the various possible corrections to the electron energy spectrum 
from beyond the Standard Model charged current interactions are studied 
in an often considered effective operator approach. The distortion 
of the spectrum in case such operators are present is analyzed. 
We summarize our results in section \ref{sec:concl}.

%% file: relativistic.tex
\section{\label{sec:rel}Fully relativistic treatment of beta decay}

We consider in this section the $\beta$-decay of a mother nucleus $\mathcal{A}$
to a daughter nucleus $\mathcal{B}$, an electron $e^{-}$ and
an electron antineutrino $\overline{\nu_e}$:
\begin{equation}\label{decay}
 \mathcal{A} \rightarrow \mathcal{B} + e^- + \overline{\nu_e}.
\end{equation}
The final electron antineutrino state $|\overline{\nu_e}\rangle$
is a superposition of mass eigenstates $|\overline{\nu_j}\rangle$. 
We will assume that apart from the three active neutrinos additional, 
necessarily sterile, neutrino species are present, \textit{i.e.}\
\begin{equation}\label{eq:2}
|\overline{\nu_e}\rangle = \sum_{j=1}^{3+n_s} U_{ej} |\overline{\nu_j}\rangle,
\end{equation}
where $n_s$ is the number of sterile neutrinos and $U$ denotes 
the $(3+n_s) \times (3+n_s)$ lepton mixing matrix. 
We will now work out general expressions for the 
electron energy spectrum, assuming only that the process in 
equation (\ref{decay}) is generated by an interaction that is mediated by particles 
much heavier than the nuclear scale. We make in this section 
no assumption about 
the Lorentz structure of the interactions. 

\subsection{Kinematics}
Our treatment of the kinematics of beta decay follows refs.~\cite{Simkovic,Masood}.
The differential decay rate of $\mathcal{A}$ is given by the sum
\begin{equation}\label{total_rate}
d\Gamma_{\mathcal{A} \rightarrow \mathcal{B} + e^- + \overline{\nu_e}} = \sum_{j=1}^{3+n_s}\hspace{0mm}
d\Gamma_{\mathcal{A} \rightarrow \mathcal{B} + e^- + \overline{\nu_j}} \, \Theta(m_{\mathcal{A}}-m_{\mathcal{B}}-m_e-m_j),
\end{equation}
where $m_j$ denotes the mass of the neutrino $\nu_j$. The $\Theta$-function has to be introduced in order
to exclude kinematically forbidden decays.

We now consider an individual term $d\Gamma_{\mathcal{A} \rightarrow \mathcal{B} + e^- + \overline{\nu_j}}$
in equation~(\ref{total_rate}):
\begin{equation}
\begin{split}
d\Gamma_{\mathcal{A} \rightarrow \mathcal{B} + e^- + \overline{\nu_j}} =
 & \frac{1}{2m_\mathcal{A}}
   \frac{d^3 p_e}{(2\pi)^3 \, 2E_e}
   \frac{d^3 p_j}{(2\pi)^3 \, 2E_j}
   \frac{d^3 p_\mathcal{B}}{(2\pi)^3 \, 2E_\mathcal{B}} \times \\
 & \left|\mathcal{M}(\mathcal{A} \rightarrow \mathcal{B} + e^- + \overline{\nu_j})\right|^2
   (2\pi)^4 \, \delta^{(4)}(p_\mathcal{A}-p_\mathcal{B}-p_e-p_j).
\end{split}
\end{equation}
Since we are interested in the electron energy spectrum only and we assume
the decaying nucleus to be unpolarized, 
$\left|\mathcal{M}(\mathcal{A} \rightarrow \mathcal{B} + e^- + \overline{\nu_j})\right|^2$
is the matrix element squared averaged over the spin of $\mathcal{A}$ and summed
over the spins of the final particles.\footnote{In the entire paper the expression $|\mathcal{M}|^2$
always means $1/2 \sum_\mathrm{spins} |\mathcal{M}(\mathrm{spins})|^2.$} This matrix element squared is Lorentz invariant
and does not depend on the spins of the particles.
Consequently, it is a function only of the masses, scalar
products $(p\cdot p')$ of $4$-momenta of the involved particles and objects of the
form
\begin{equation}\label{epsilon-inv}
\epsilon^{\mu\nu\rho\sigma} p_\mu p'_\nu p''_\rho p'''_\sigma.
\end{equation}
However, due to energy-momentum conservation there are only three independent
4-momenta in the process. Thus the expression of equation~(\ref{epsilon-inv})
always vanishes due to the total antisymmetry of the $\epsilon$-symbol.

Taking into account only (possibly effective) tree-level contributions,
the amplitude $\mathcal{M}$
has the form
\begin{equation}
\mathcal{M} = [\overline{u}_e \mathcal{O} v_j]
[\overline{u}_\mathcal{B} \mathcal{O}' u_\mathcal{A}],
\end{equation}
where $\mathcal{O}$ and $\mathcal{O}'$ are $4\times 4$-matrices.
If $\mathcal{O}$ and $\mathcal{O}'$ do not depend on the 4-momenta
of the particles, the only source of 4-momenta are the four spinors
$\overline{u}_e,\, v_j,\, \overline{u}_\mathcal{B},\, u_\mathcal{A}$.
In this case $|\mathcal{M}|^2$ is a quadratic polynomial in products of 4-momenta,
\textit{i.e.}\ all momentum-dependent terms of $|\mathcal{M}|^2$
are of the form
\begin{equation}
(p\cdot p') \quad\text{or}\quad (p\cdot p')(p''\cdot p''').
\end{equation}
In this paper the only case of a momentum-dependent matrix $\mathcal{O}$
or $\mathcal{O}'$ will be a (weak magnetism) contribution to $\mathcal{O}'$
proportional to 
$q/M$,
with $q=p_\mathcal{A}-p_\mathcal{B}$ being the momentum transfer and
$M_N$ being a mass scale of the order of the nucleus mass.
This will induce terms in $|\mathcal{M}|^2$ of the form
\begin{equation}\label{sixmomenta}
\frac{(p \cdot q)(p'\cdot q)(p'' \cdot p''')}{M_N^2}
\quad
\text{or}
\quad
\frac{(p \cdot p')(p''\cdot p''')(q\cdot q)}{M_N^2}.
\end{equation}
Since we will focus on tritium decay for which $q < 20\,\text{keV}$
and $M_N \sim 3\,\text{GeV}$, these contributions are suppressed by a
factor of $(q/M_N)^2 \lesssim 10^{-10}$ and are therefore negligible.
Thus, with excellent accuracy, $|\mathcal{M}|^2$ is a quadratic polynomial
in products of the form $p\cdot p'$.
However, due to energy-momentum conservation
\begin{equation}
p_\mathcal{A} = p_\mathcal{B} + p_e + p_j
\end{equation}
only two products of $4$-momenta
are independent. For our purposes it will be most convenient
to express all products of $4$-momenta in terms of
\begin{equation}
p_\mathcal{A}\cdot p_{e} \quad\text{and}\quad p_\mathcal{A}\cdot p_j.
\end{equation}
Since the decay rate is defined in the rest frame of the decaying particle
we find
\begin{equation}
p_\mathcal{A}\cdot p_{e} = m_\mathcal{A}E_e \quad\text{and}\quad p_\mathcal{A}\cdot p_j = m_\mathcal{A}E_j.
\end{equation}
Thus, the matrix element squared is a function of the particle masses and
the electron and neutrino energy only. As discussed above, this function must be
a quadratic polynomial in $E_e$ and $E_j$,
\textit{i.e.}\
\begin{equation}\label{parameterization}
\left|\mathcal{M}(\mathcal{A} \rightarrow \mathcal{B} + e^- + \overline{\nu_j})\right|^2 =
A + B_1 E_e + B_2 E_j + C E_e E_j + D_1 E_e^2 + D_2 E_j^2,
\end{equation}
where $A,\,B_1,\,B_2,\,C,\,D_1$ and $D_2$ are functions of the particle masses 
and coupling constants.
Since $|\mathcal{M}|^2$ therefore does not depend on the
direction of the emitted electron, the computation of $d\Gamma/dE_e$
is possible without knowledge of the explicit form of $|\mathcal{M}|^2$---see
appendix~\ref{appA}.
The result of this computation is
\begin{equation}
 \left( \frac{d\Gamma}{dE_e} \right)_{\overline{\nu_j}} = \frac{1}{64\pi^3 m_\mathcal{A}}
 \int_{E_{j-}}^{E_{j+}} dE_j
 \left|\mathcal{M}(\mathcal{A} \rightarrow \mathcal{B} + e^- + \overline{\nu_j})\right|^2,
\end{equation}
where
\begin{equation}
E_{j\pm} = \frac{-(m_\mathcal{A}-E_e)(E_e m_\mathcal{A}-\alpha) \pm |\vec{p}_e|
\sqrt{(E_e m_\mathcal{A} -\alpha + m_j^2)^2 -m_\mathcal{B}^2 m_j^2}}{m_\mathcal{A}^2-2m_\mathcal{A}E_e + m_e^2}
\end{equation}
and
\begin{equation}
\alpha = \frac{1}{2} \left( m_\mathcal{A}^2 - m_\mathcal{B}^2 + m_e^2 + m_j^2 \right).
\end{equation}
With $|\mathcal{M}|^2$ given by equation~(\ref{parameterization}) the
$E_j$-integration is trivial and gives
\begin{equation}\label{main}
 \begin{split}
 & \left( \frac{d\Gamma}{dE_e} \right)_{\overline{\nu_j}} = 
  \frac{1}{64\pi^3 m_\mathcal{A}} \times \\
 & \left\{
 (A + B_1 E_e + D_1 E_e^2) (E_{j+} - E_{j-}) +
 \frac{1}{2} (B_2 + C E_e) (E_{j+}^2 - E_{j-}^2) +
 \frac{1}{3} D_2 (E_{j+}^3 - E_{j-}^3) \right\}.
 \end{split}
\end{equation}
Equations (\ref{parameterization}) and (\ref{main}) define the most general 
electron energy spectrum in $\beta$-decay. Any charged current interaction 
will specify the functions $A, B_{1,2}, C, D_{1,2}$, which can then be 
inserted in those expressions. For the Standard Model the 
result is presented in equation (\ref{SM_para}), and the various 
possible new physics cases are treated in section \ref{sec:NP}.

Defining
\begin{equation}\label{PQ}
P(E_e) \equiv -\frac{(m_\mathcal{A}-E_e)(E_e m_\mathcal{A}-\alpha)}{m_\mathcal{A}^2-2m_\mathcal{A}E_e + m_e^2},\quad
Q(E_e) \equiv \frac{ |\vec{p}_e|
\sqrt{(E_e m_\mathcal{A} -\alpha + m_j^2)^2 -m_\mathcal{B}^2 m_j^2}}{m_\mathcal{A}^2-2m_\mathcal{A}E_e + m_e^2},
\end{equation}
we have $E_{j\pm} = P\pm Q$ and thus
\begin{equation}\label{relativistic_spectrum}
 \left( \frac{d\Gamma}{dE_e} \right)_{\overline{\nu_j}} = 
  \frac{Q(E_e)}{32\pi^3 m_\mathcal{A}} \times \\
 \left\{
 (A + B_1 E_e + D_1 E_e^2) +
 (B_2 + C E_e) P(E_e) +
 D_2 ( P^2(E_e) + \frac{1}{3} Q^2(E_e) ) \right\}.
\end{equation}
Note that almost all quantities in the above equation depend on the
neutrino mass and are therefore different for different contributions to the
total decay rate. In particular also
the maximal electron energy
\begin{equation}
E_e^\text{max} =
\frac{m_\mathcal{A}^2 + m_e^2 - (m_\mathcal{B} + m_j)^2}{2 m_\mathcal{A}}
\end{equation}
is a function
of the neutrino mass.
As pointed out in~\cite{Masood:2005aj}
the difference to the usual non-relativistic approximation
$(E_e^\text{max})_\text{NR} \equiv \mA-\mB-m_j$ can be substantial.
For tritium decay and low neutrino masses $\lesssim 10 \,\text{eV}$ one obtains
$(E_e^\text{max})_\text{NR}-E_e^\text{max} \approx
3.4\,\text{eV}$~\cite{Masood:2005aj}. For a neutrino mass of
$5\,(10, \, 15)\, \text{keV}$, the difference is about
$2.5\,(1.6,\,0.6) \,\text{eV}$. 

Taking into account that the decay is kinematically forbidden if
$m_j > \mA - \mB - m_e$ and that $\left( \frac{d\Gamma}{dE_e} \right)_{\overline{\nu_j}}$
contributes only for $E_e < E_e^\text{max}(m_j)$,
we find the total electron spectrum:
\begin{equation}\label{fullspectrum}
\frac{d\Gamma}{dE_e} = \sum_{j=1}^{3+n_s}
 \left( \frac{d\Gamma}{dE_e} \right)_{\overline{\nu_j}}
 \Theta(E_e^\text{max}(m_j) - E_e)
 \Theta(\mA - \mB - m_e - m_j).
\end{equation}
In the following, we will
use the abbreviation
\begin{equation}
\widetilde\Theta_j \equiv \Theta(E_e^\text{max}(m_j) - E_e) \Theta(\mA - \mB - m_e - m_j).
\end{equation}

Since each term in the sum~(\ref{fullspectrum}) is proportional to
$|\vec{p}_e|$ (recall that the spectrum from equation (\ref{relativistic_spectrum})
is proportional to $Q(E_e) \propto |\vec{p}_e|$), the whole spectrum is 
proportional to $|\vec{p}_e|$,
and in particular
\begin{equation}
\frac{d\Gamma}{dE_e}\Big\vert_{E_e=m_e} = 0.
\end{equation}
Furthermore, since $Q(E_e^\text{max})=0$, see appendix~\ref{appB}, the endpoint of the spectrum
is reached at
\begin{equation}
E_e^\text{end} =  \frac{m_\mathcal{A}^2 + m_e^2 - (m_\mathcal{B} + m_0)^2}{2 m_\mathcal{A}},
\end{equation}
where $m_0$ is the mass of the lightest neutrino mass eigenstate. Thus, the shift
of the endpoint compared to the case of at least one massless neutrino is given by
\begin{equation}
E_e^\text{end}\Big\vert_{m_0=0} - E_e^\text{end} = \frac{2 \mB m_0 + m_0^2}{2 \mA} \approx \frac{\mB}{\mA} m_0 \approx m_0.
\end{equation}
The difference of the exact expression to the approximate 
expression $m_0$ is $-1.9\cdot 10^{-5}$ eV for $m_0 = 0.1$ eV, 
$-1.9\cdot 10^{-4}$ eV for $m_0 = 1$ eV, $-0.19$ eV for $m_0 = 1$ keV and 
 $-0.94$ eV for $m_0 = 5$ keV.

In order to simplify the expression of the spectrum 
we expand the functions $P(E_e)$ and $Q(E_e)$ from equation 
(\ref{PQ}) 
in terms of the small parameters~\cite{Griffiths}
\begin{equation}\label{expansionpar}
\epsilon \equiv \frac{\mA-\mB}{\mA},\quad
\delta \equiv \frac{m_e}{\mA},\quad
\eta \equiv \frac{E_e}{\mA}
\quad\text{and}\quad \rho\equiv \frac{m_j}{\mA}.
\end{equation}
Taking the standard example of tritium decay,
$\mA = m(^3\mathrm{H}^+)$, $\mB = m(^3\mathrm{He}^{2+})$---see
table~\ref{masses}---we find
\begin{equation}
\eta < \epsilon = 1.9 \times 10^{-4},\quad
\delta = 1.8 \times 10^{-4},\quad \rho< \epsilon - \delta = 6.7 \times 10^{-6},
\end{equation}
\textit{i.e.}\ all expansion parameters are smaller than $2\times 10^{-4}$.
The parameter $\rho$ (even for large neutrino masses $\sim\text{keV}$) is
smaller by at least one order of
magnitude.\footnote{If one would actually use the expansion in the parameters
of equation~(\ref{expansionpar}) for numerical estimations---which we will not
do here---one has to keep in mind that for $m_j\lesssim \mathrm{eV}$, $\rho$
may be smaller or of comparable size to $\eta^2$, $\epsilon^2$ and $\delta^2$,
in which case an expansion to second order in the small parameters may be
necessary to estimate the effect of nonvanishing neutrino masses on the spectrum.}
The reason for this is the small energy
release $\mA-\mB-m_e \lesssim 18.591~\text{keV}$ of tritium beta decay compared to the mass
of the mother nucleus.

We expand the functions $P(E_e)$ and $Q(E_e)$ to lowest order in terms of the four
expansion parameters of equation~(\ref{expansionpar}). For this purpose we
treat each of the parameters as being of the same order $\lambda$,
\textit{i.e.}\ $\epsilon \sim \delta \sim \eta \sim \rho \sim \lambda$.
The results are shown in table~\ref{expansion}. From there we find
that $Q(E_e)$ is suppressed with respect to $P(E_e)$ by a factor
of
\begin{equation}
\frac{|\vec{p}_e|}{\mA} < \frac{\mA - \mB - m_e}{\mA} = \epsilon - \delta = 6.7\times 10^{-6},
\end{equation}
the numeric value being again for tritium decay. Consequently,
\begin{equation}
Q \ll P \quad \text{and} \quad Q^3 \lll QP^2.
\end{equation}
These inequalities hold also close to the endpoint of the spectrum where
\begin{equation}
Q(E_e^\text{max}) = 0,\quad P(E_e^\text{max}) = \frac{m_j \left( \mA^2 + (\mB + m_j)^2 -m_e^2 \right) }{2\mA (\mB + m_j)}
\end{equation}
and
\begin{equation}
\lim_{E_e\rightarrow E_e^\text{max}} \frac{Q(E_e)}{P(E_e)} = 0.
\end{equation}
From the lowest order expansion in table~\ref{expansion}, setting $m_j=0$
we find the approximate properties
\begin{equation}
 P(E_e) \propto (\mA-\mB-E_e)
\quad\text{and}\quad
 Q(E_e) \propto |\vec{p}_e| (\mA-\mB-E_e),
\end{equation}
\textit{i.e.}\ $\!P(E_e)$ is a linear function of $E_e$, and $Q(E_e)$ is proportional
to the product of this linear function with $|\vec{p}_e|$.

\begin{table}
\begin{center}
\renewcommand{\arraystretch}{1.4}
\begin{tabular}{|l|l|l|}
\hline
function & lowest order expansion & for $m_j=0$ \\
\hline
$Q(E_e)$ & $\mA \frac{|\vec{p}_e|}{\mA} \left( \sqrt{(\epsilon -\eta)^2 -\rho^2} + \mathcal{O}(\lambda^2)\right)$ &
$\mA \frac{|\vec{p}_e|}{\mA} \left( \epsilon -\eta + \mathcal{O}(\lambda^2)\right)$ \\
$P(E_e)$ & $\mA \left( \epsilon -\eta + \mathcal{O}(\lambda^2)\right)$ &
$\mA \left( \epsilon -\eta + \mathcal{O}(\lambda^2)\right)$ \\
$Q(E_e)P(E_e)$ & $\mA^2 \frac{|\vec{p}_e|}{\mA} \left( (\epsilon-\eta) \sqrt{(\epsilon -\eta)^2 -\rho^2} + \mathcal{O}(\lambda^3)\right)$ &
$\mA^2 \frac{|\vec{p}_e|}{\mA} \left( (\epsilon-\eta)^2 + \mathcal{O}(\lambda^3)\right)$ \\
$Q(E_e)P(E_e)^2$ & $\mA^3 \frac{|\vec{p}_e|}{\mA} \left( (\epsilon-\eta)^2 \sqrt{(\epsilon -\eta)^2 -\rho^2} + \mathcal{O}(\lambda^4)\right)$ &
$\mA^3 \frac{|\vec{p}_e|}{\mA} \left( (\epsilon-\eta)^3 + \mathcal{O}(\lambda^4)\right)$ \\
$Q(E_e)^3$ & $\mA^3 \left(\frac{|\vec{p}_e|}{\mA}\right)^3 \left( ((\epsilon -\eta)^2 -\rho^2)^{3/2} + \mathcal{O}(\lambda^4)\right)$ &
$\mA^3 \left(\frac{|\vec{p}_e|}{\mA}\right)^3 \left( (\epsilon -\eta)^3 + \mathcal{O}(\lambda^4)\right)$ \\
\hline
\end{tabular}
\renewcommand{\arraystretch}{1.0}
\caption{Expansion of the functions $P(E_e)$ and $Q(E_e)$ and their products
in terms of the small parameters $\epsilon \sim \delta \sim \eta \sim \rho \sim \lambda$.}\label{expansion}
\end{center}
\end{table}

%% file: SM.tex
\section{\label{sec:SM}The electron spectrum of tritium beta decay in the Standard Model}

Let us now apply the formalism from the previous section in detail 
to the $\beta$-decay of tritium.
Moreover, taking tritium decay as an example,
we will make the transition from the general spectrum to the well-known textbook results.

\subsection{Shape of the spectrum and corrections to the non-relativistic case}

\begin{figure}
\begin{center}
\includegraphics[width=0.5\textwidth]{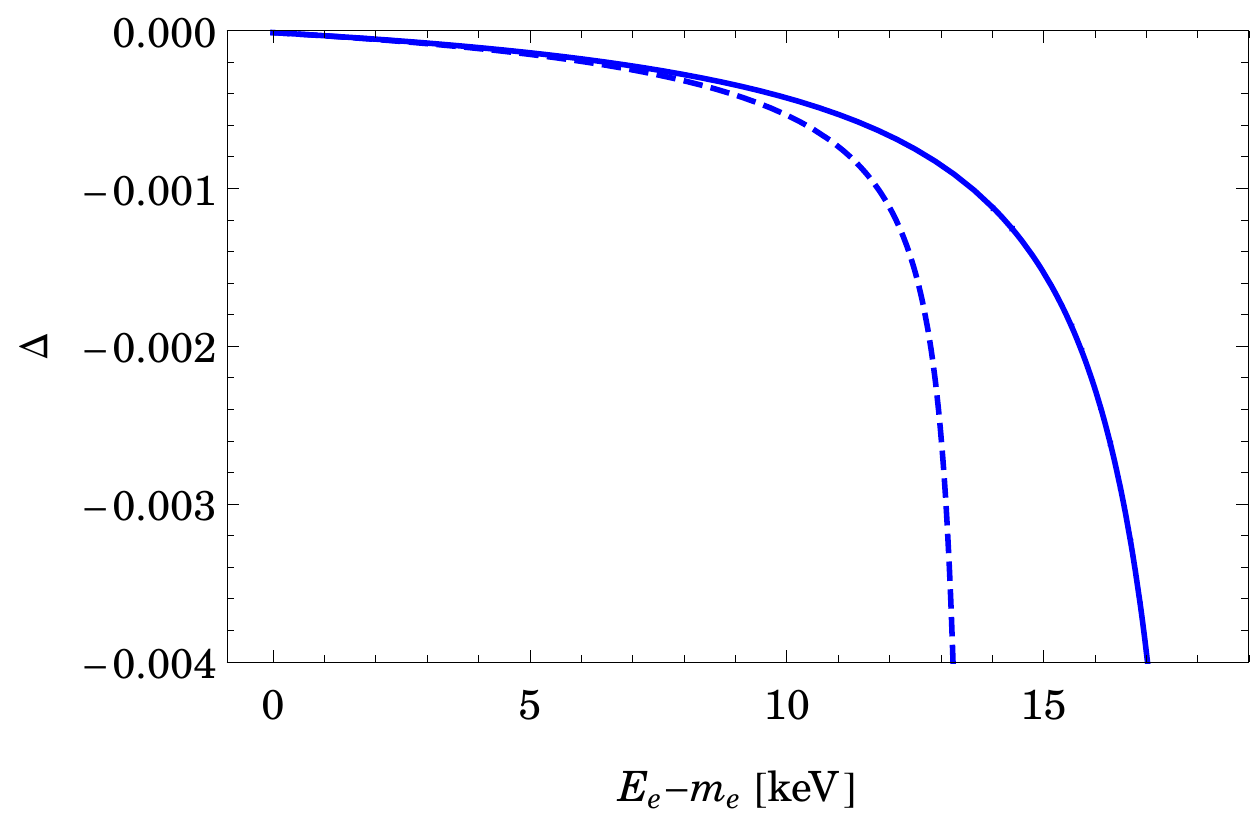}\hspace*{5mm}
\includegraphics[width=0.5\textwidth]{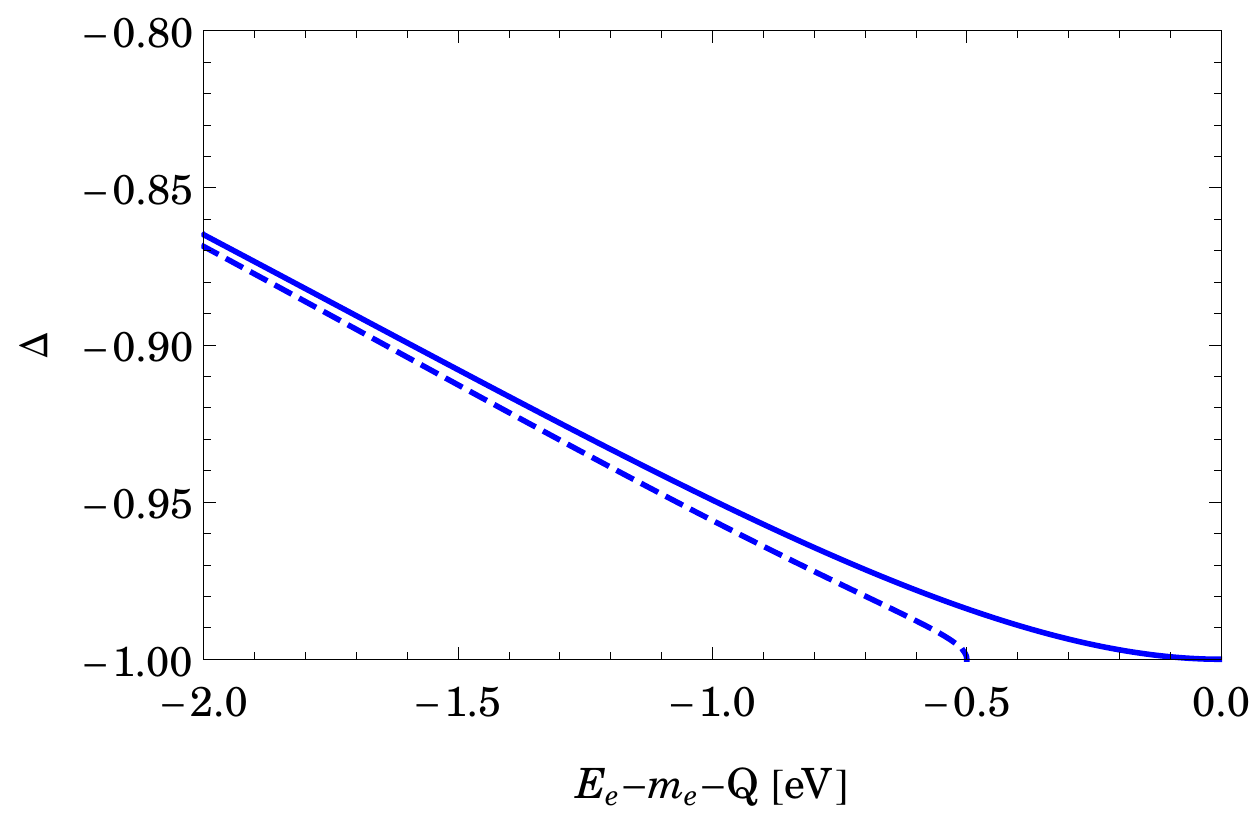}\hspace*{5mm}
\end{center}
\caption{Left: Plot of the relative deviation $\Delta$ (see equation (\ref{Delta}))
between the non-relativistic and relativistic result for the beta decay of tritium
assuming only one single neutrino species with $|U_{ej}|=1$ and a 
mass of $m_j=0$ (solid line) and $m_j=5\,\mathrm{keV}$ (dashed line). 
Right: The same plot for the region near the endpoint for the values 
$|U_{ej}|=1$, $m_j=0$ (solid line) and $m_j=0.5 \,\mathrm{eV}$ (dashed line).}
\label{figures-deviation}
\end{figure}

Assuming the nuclei to be point particles interacting only via the
weak interaction, the Standard Model effective Lagrangian for the
$\beta$-decay
$ \mathcal{A} \rightarrow \mathcal{B} + e^- + \overline{\nu_e}$ 
is given by
\begin{equation}\label{SM-Lagrangian-simple}
-\frac{G_F}{\sqrt{2}} V_{ud}
\left( \overline{e} \gamma^\mu (\mathbbm{1}-\gamma^5) \nu_e \right)
\left( \overline{\mathcal{B}} \gamma_\mu ( g_V \mathbbm{1}- g_A \gamma^5) \mathcal{A} \right) + \mathrm{H.c.}
\end{equation}
with $\mathcal{A}=\hspace{0mm}^3\mathrm{H}^+$ and
$\mathcal{B}=\hspace{0mm}^3\mathrm{He}^{2+}$. 
Here 
we use 
the elementary particle treatment
of weak processes~\cite{Kim-Primakoff1,Kim-Primakoff2,Kim} as applied
to tritium beta decay in~\cite{Wu-Repko,Simkovic}, 
\textit{i.e.}\ we use the fact that the transition $^3\mathrm{H}^+\rightarrow \hspace{0mm}^3\mathrm{He}^{2+} + e^- + \overline{\nu}_e$
has the same relevant spin and isospin structure as neutron decay
$n\rightarrow p + e^- + \overline{\nu}_e$. 
In this case, in the first approximation, the effects from nuclear physics\footnote{For a more detailed
discussion of nuclear effects see section~\ref{sec:SMcor}.} can be
absorbed into two form factors $g_V$ and $g_A$. 
At tree-level the matrix element squared, already averaged over the spin orientations of $\mathcal{A}$
and summed over the spins of the final state particles\footnote{Both
$^3\mathrm{H}^+$ and $^3\mathrm{He}^{2+}$ have spin 1/2.}
is given by
\begin{equation}\label{SM-matrix-element}
\begin{split}
\left|\mathcal{M}(\mathcal{A} \rightarrow \mathcal{B} + e^- + \overline{\nu_j})\right|^2 = \enspace
& 16\, G_F^2\, |V_{ud}|^2\, |U_{ej}|^2 \times \\
& \Big\{ (g_V+g_A)^2\, (p_j \cdot p_\mathcal{A})\,(p_e \cdot p_\mathcal{B}) + \\
& \phantom{\Big\{}  (g_V-g_A)^2\, (p_j \cdot p_\mathcal{B})\,(p_e \cdot p_\mathcal{A}) - \\
& \phantom{\Big\{}  (g_V^2-g_A^2)\, m_\mathcal{A} m_\mathcal{B}\,(p_j \cdot p_e) \Big\}.
\end{split}
\end{equation}
Using energy-momentum conservation we can reformulate this as a polynomial in $E_e$ and $E_j$
with the coefficients\footnote{Actually, including the weak magnetism
correction to be discussed in section \ref{sec:SMcor} induces a
small contribution to $C$ and corrections to the other
parameters. The overall effect on the total decay width is
$1.8 \times 10^{-4}\,\%$.}
\begin{subequations}\label{SM_para}
\begin{align}
& A = \frac{\gamma}{2} m_\mathcal{A} m_\mathcal{B}\, (g_V^2 - g_A^2)  (m_{\mathcal{A}}^2 - m_\mathcal{B}^2 + m_e^2 + m_j^2),\\
& B_1 = \frac{\gamma}{2} \mA\, \left\{ (g_V - g_A)^2 ( \mA^2 - \mB^2 + m_e^2 - m_j^2 ) - 2 \mA\mB (g_V^2-g_A^2) \right\},\\
& B_2 = \frac{\gamma}{2} \mA\, \left\{ (g_V + g_A)^2 ( \mA^2 - \mB^2 - m_e^2 + m_j^2 ) - 2 \mA\mB (g_V^2-g_A^2) \right\},\\
& C = 0,\\
& D_1 = -\gamma \mA^2 (g_V-g_A)^2,\\
& D_2 = -\gamma \mA^2 (g_V+g_A)^2,
\end{align}
\end{subequations}
where we have defined the overall constant
\begin{equation}
 \gamma \equiv 16\, G_F^2\, |V_{ud}|^2\, |U_{ej}|^2.
\end{equation}
From these results we can recover the ``classic textbook result'' by setting
$g_V=g_A=1$ which gives
\begin{equation}\label{simplestSM_par}
A=B_1=C=D_1=0
\quad\text{and}\quad
B_2 = 2\gamma \mA \left( \mA^2 - \mB^2 -m_e^2 +m_j^2 \right),\,
D_2 = -4\gamma \mA^2.
\end{equation}
Using the general spectrum from equation (\ref{relativistic_spectrum}) and the expansion parameters 
defined in (\ref{expansionpar}), the electron energy spectrum to lowest order is given by
\begin{equation}
\begin{split}
\left( \frac{d\Gamma}{dE_e} \right)_{\overline{\nu}_j}
& = \frac{\gamma \mA^4}{8\pi^3} \, \sqrt{\eta^2-\delta^2}\, \eta\, (\epsilon-\eta)^2
\sqrt{1-\left(\frac{\rho}{\epsilon-\eta}\right)^2}
+ \mathcal{O}(\lambda^5)\\
& = \frac{\gamma \mA^3}{8\pi^3} \, |\vec{p}_e|\, \eta\, (\epsilon-\eta)^2
\sqrt{1-\left(\frac{\rho}{\epsilon-\eta}\right)^2}
+ \mathcal{O}(\lambda^5)\\
& = \frac{2}{\pi^3} G_F^2 |V_{ud}|^2 |U_{ej}|^2 \, |\vec{p}_e| E_e (\mA-\mB-E_e)^2
\sqrt{1-\left(\frac{\rho}{\epsilon-\eta}\right)^2}
+ \mathcal{O}(\lambda^5).
\end{split}
\end{equation}
Summing over the three neutrino species we obtain
\begin{equation}\label{SM_LO}
\begin{split}
\frac{d\Gamma}{dE_e} = \sum_j \left( \frac{d\Gamma}{dE_e} \right)_{\overline{\nu}_j}
\widetilde{\Theta}_j = \; &
\frac{2\, G_F^2\, |V_{ud}|^2}{\pi^3} \, |\vec{p}_e| E_e (\mA-\mB-E_e)^2 \times \\
& \times \left(\sum_{j} |U_{ej}|^2 \sqrt{1-\frac{m_j^2}{(\mA-\mB-E_e)^2}} \widetilde{\Theta}_j \right)
+ \mathcal{O}(\lambda^5) \\
& \equiv \left(\frac{d\Gamma}{dE_e}\right)_\text{NR} + \mathcal{O}(\lambda^5)\,
.
\end{split}
\end{equation}
Here we have defined the lowest-order approximation $(\frac{d\Gamma}{dE_e})_\text{NR}$, 
which is of order $\lambda^4$. 
Since the expansion in $\lambda$
corresponds to an expansion in $1/\mA$, the lowest order
term in~(\ref{SM_LO}) is the result to be expected from a non-relativistic
computation.
Indeed, setting the neutrino masses to zero one obtains
\begin{equation}\label{SM_NR}
\left(\frac{d\Gamma}{dE_e}\right)_{\text{NR},\, m_j=0} =
\frac{2\, G_F^2\, |V_{ud}|^2}{\pi^3} \, |\vec{p}_e| E_e (\mA-\mB-E_e)^2
\widetilde{\Theta}_j\vert_{m_j=0}
\end{equation}
which is the classic non-relativistic textbook
result~\cite{Griffiths}. 
We now want to compare the exact relativistic spectrum obtained using~(\ref{relativistic_spectrum})
to the non-relativistic approximation of equation~(\ref{SM_LO})
by studying the relative deviation
\begin{equation}\label{Delta}
\Delta \equiv 
\frac{\left(d\Gamma/dE_e\right)-
\left(d\Gamma/dE_e\right)_{\text{NR}}}{\left(d\Gamma/dE_e\right)_{\text{NR}}}.
\end{equation}
For simplicity we assume only one neutrino with $|U_{ej}|=1$.
Expanding in $\lambda$ one obtains $\Delta=\mathcal{O}(\lambda)$,
\textit{i.e.}\ $|\Delta|\sim 10^{-4}\div 10^{-3}$ for tritium decay.
In figure~\ref{figures-deviation} the quantity $\Delta$ is plotted for a massless
and a keV neutrino
for the spectrum following from the parameters of
equation~(\ref{simplestSM_par}).
Indeed, not too close to the endpoint, we numerically find
$|\Delta|\sim 10^{-4}\div 10^{-3}$.
Approaching the endpoint, $\Delta$ goes to $-1$.
The reason for this is that
\begin{equation}
\lim_{E_e \rightarrow E_e^\text{max}} \left(\frac{d\Gamma}{dE_e}\right)_{\text{NR}} \neq 0,
\end{equation}
while the exact spectrum of course vanishes at the endpoint.

Let us finally comment on the applicability of the results obtained so
far, which may be estimated most easily by computing the half-life of
tritium using our relativistic Standard Model expression for $d\Gamma/dE_e$.
For the computation we use $m_j=0$, and the 
experimental data of table~\ref{masses}, in particular we take $g_A = 1.2646$. 
Naively inserting numbers we find $t_{1/2} \approx 17.1\,\text{yr}$, which is
$40\,\%$ larger than the value for $^3\mathrm{H}^+$-decay estimated from
experiment
$t_{1/2}(^3\mathrm{H}^+) = (12.238\pm0.020)\,\text{yr}$~\cite{triton-decay}.
The main reason for this deviation is our ignoring of the
electromagnetic interaction between the newly formed $^3\mathrm{He}^{2+}$-nucleus
and the emitted electron. This can be taken into account by multiplying
$d\Gamma/dE_e$ with the Fermi function $F(Z,E_e)$~\cite{Fermi},
\textit{i.e.}\
\begin{equation}
 \frac{d\Gamma}{dE_e} \rightarrow  \frac{d\Gamma}{dE_e} F(Z,E_e).
\end{equation}
In units where $\hbar=c=1$ the Fermi function is given by~\cite{Konopinski}
\begin{equation}
F(Z,E_e) = 2(1+\gamma) (2pR)^{-2(1-\gamma)} e^{\pi y}
\frac{|\Gamma(\gamma+iy)|^2}{\Gamma(2\gamma+1)^2},
\end{equation}
where $\Gamma$ here denotes the gamma function and
\begin{equation}
 \gamma = (1-\alpha_\text{EM}^2 Z^2)^{1/2},~ 
 y = \alpha_\text{EM} Z E_e/p_e.
\end{equation}
The atomic number of the daughter nucleus $Z$ is 2 for tritium decay, 
$\alpha_\text{EM}$ is the electromagnetic fine structure constant
and $R$ is the radius of the daughter nucleus. One can conveniently
express $R$ in units of $m_e^{-1}$. We will adopt the value
$R=2.8840 \times 10^{-3}\,m_e^{-1}$
used in~\cite{tritium-sensitivity} for $^3\hspace{0mm}\mathrm{He}$.

Including the Fermi function we find a half-life of $11.9\,\text{yr}$,
which is off the experimental value by less than $3\%$. Therefore,
the electromagnetic interaction makes up a substantial part of
the decay rate of tritium.\footnote{Including the 
weak magnetism term has an effect of only $10^{-4}\,\%$ on the half-life.} 
However, there are many other effects to be taken into account
aiming at interpretation of high-precision measurements of 
$d\Gamma/dE_e$. We will discuss all these effects 
in section~\ref{sec:SMcor}. 

\subsection{\label{sec:Kurie}Kurie plots and the endpoint of the spectrum}

\begin{figure}
\begin{center}
\includegraphics[width=0.45\textwidth]{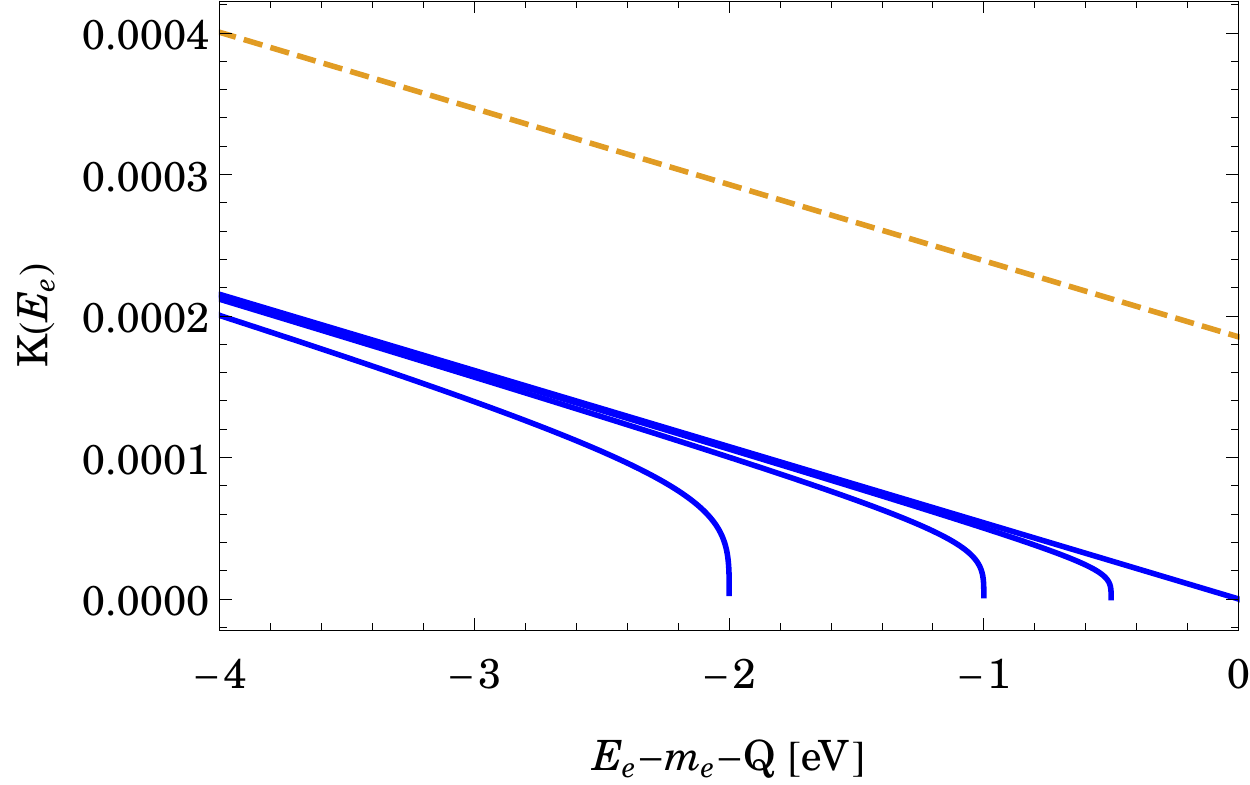}
\quad
\includegraphics[width=0.425\textwidth]{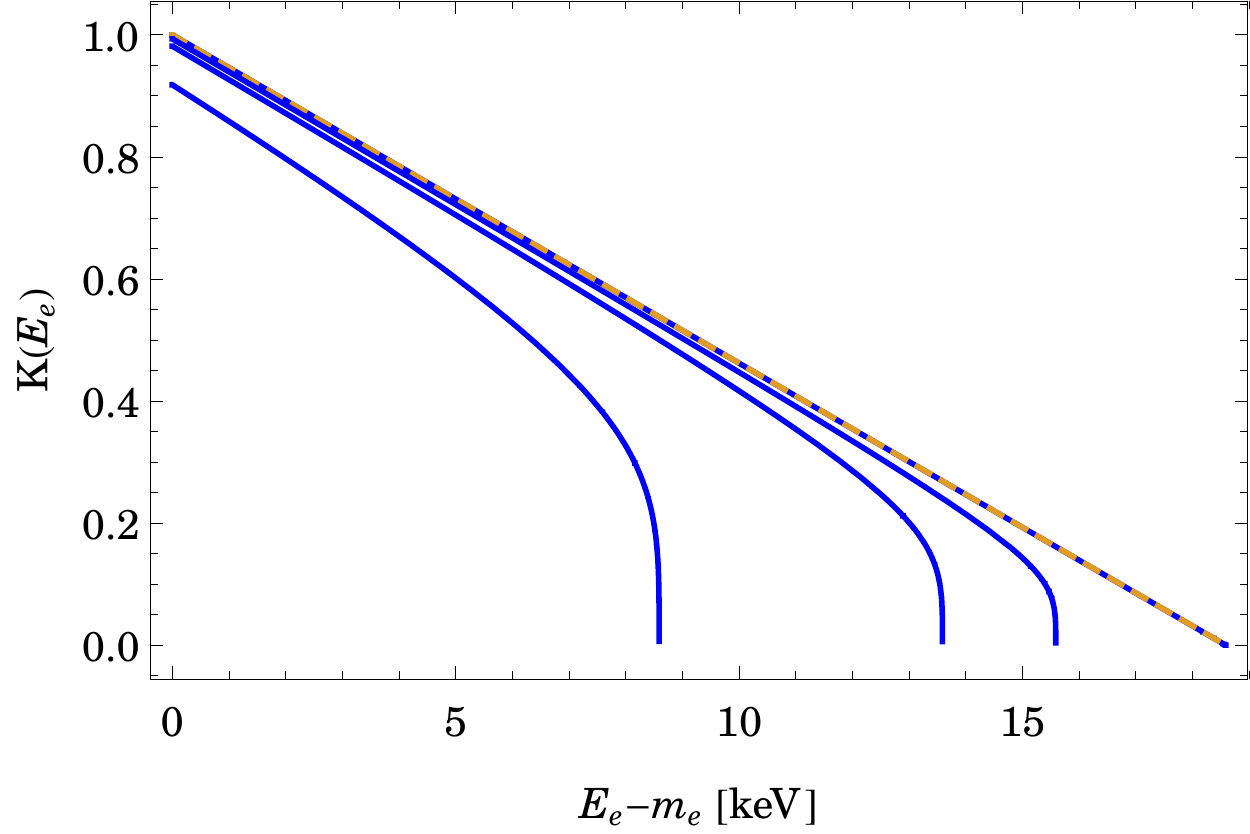}
\caption{Plots of the Kurie-like function $K(E_e)$ (see equation (\ref{Kurie}))
for the spectrum following from the parameters of
equation~(\ref{simplestSM_par}),
for simplicity assuming
only one neutrino species.
Left plot:
The solid lines correspond to (from left to right) $m_j=2.0,\,1.0,\,0.5$ and $0$ eV, respectively.
The dashed curve is the non-relativistic approximation~(\ref{linear_Kurie}) for $m_j=0$. As can be seen
from this plot, the effect of neglecting the relativistic corrections to $K(E_e)$ is, in absolute
numbers, much larger than the effect of a nonvanishing neutrino mass at the level of $m_j<1\,\text{eV}$.
The reason for this is the large difference
$(E_e^\text{max})_\text{NR}-E_e^\text{max} = 3.4\,\text{eV}$. 
Right plot: The same plot for neutrino masses (from left to right) of 10, 5, 3 and 0~keV.}
\label{Kurie-plots}
\end{center}
\end{figure}

The effect of 
non-zero 
neutrino masses on the spectral endpoint can
be seen best in plots of the Kurie-like function
\begin{equation}\label{Kurie}
K(E_e) \equiv \frac{1}{\mA - \mB} \sqrt{\frac{d\Gamma/dE_e}{G_0(E_e)}},
\end{equation}
where
\begin{equation}
G_0(E_e) \equiv \frac{2 G_F^2 |V_{ud}|^2}{\pi^3} |\vec{p}_e| E_e \, F(Z,\,E_e).
\end{equation}
The lowest order approximation of $K(E_e)$ in the Standard Model
for $g_V=g_A=1$ and $m_j=0$---see equation~(\ref{SM_NR})---is then given by the linear function
\begin{equation}\label{linear_Kurie}
K(E_e) = 1-\frac{E_e}{\mA-\mB}.
\end{equation}
Assuming non-zero neutrino masses or including the terms of
$\mathcal{O}(\lambda^5)$ of 
equation~(\ref{SM_LO}) 
will lead to
deviations from~(\ref{linear_Kurie}). 
The endpoints of the Kurie plots for different values of $m_j$
are shown in figure~\ref{Kurie-plots}.

Before we go on to discuss corrections from Standard Model physics, let us discuss
the effect of non-vanishing neutrino masses on the shape of the endpoint of the Kurie plot.
Using the lowest-order approximation
$\left(\frac{d\Gamma}{dE_e}\right)_\text{NR}$
of equation~(\ref{SM_LO}) one finds the Kurie function
\begin{equation}
K(E_e) = \left(1-\frac{E_e}{\mA-\mB}\right) \times 
\sqrt{
\sum_{j} |U_{ej}|^2 \sqrt{1-\frac{m_j^2}{(\mA-\mB-E_e)^2}} \widetilde{\Theta}_j
},
\end{equation}
\textit{i.e.}\ the linear function of equation~(\ref{linear_Kurie})
multiplied by a correction term which goes to 1 for vanishing neutrino mass.
Far from the endpoint ($\mA-\mB-E_e \gg m_j$) we find
\begin{equation}
\sqrt{\sum_{j} |U_{ej}|^2 \sqrt{1-\frac{m_j^2}{(\mA-\mB-E_e)^2}} \widetilde{\Theta}_j} \simeq
1-\frac{\sum_j |U_{ej}|^2 m_j^2}{4(\mA-\mB-E_e)^2},
\end{equation}
\textit{i.e.}\ the deviation of the Kurie function from the
case of vanishing neutrino mass is proportional to the effective neutrino mass
squared~\cite{Shrock:1980vy}
\begin{equation}
m_\beta^2 \equiv \sum_j |U_{ej}|^2 m_j^2\,.
\end{equation}
Clearly, the effect of nonvanishing neutrino masses becomes strong
in the region where $\mA-\mB-E_e \sim m_\beta$, that is for
\begin{equation}
E_e \sim \mA-\mB-m_\beta,
\end{equation}
\textit{i.e.}\ an energy $m_\beta$ before the endpoint of the linear Kurie function.
Also other effective neutrino masses like
\begin{equation}
m_\beta' \equiv \sum_j |U_{ej}|^2 m_j
\end{equation}
have been considered in the literature~\cite{Weinheimer,Vissani,Smirnov}. However, in the range of
sensitivity of KATRIN (which would mean quasi-degenerate neutrinos),
they all coincide.

%% file: SM_corr.tex
\subsection{\label{sec:SMcor}Corrections from Standard Model physics}

Up to now we have treated the ideal case of pointlike tritium nuclei
decaying into helium-3 nuclei ignoring the electromagnetic
and strong interaction. However,
the actual experimental situation is of course much more complex.
As we already saw, the electromagnetic interaction between
the helium nucleus and the outgoing electron (taken into account by the Fermi
function $F(Z,E_e)$) is responsible for a large part of the decay rate.
Also QED radiative corrections have to be taken into account
to correctly interpret the results of high-precision measurements
of the electron spectrum.
Moreover, the source in a tritium decay experiment is not composed of tritium
nuclei, but tritium molecules in gaseous state at finite
temperature ($T=30\,\mathrm{K}$ for the KATRIN experiment~\cite{KATRIN-T}).
The theory corrections which have to be taken into account to
make interpretations of high-precision data on tritium beta
decay in terms of bounds on new physics possible at all
are summarized in~\cite{tritium-sensitivity} and include:
\begin{itemize}
 \item Excited final states: The initial state of the decay is
a tritium molecule $^3\mathrm{H}_2$. However, the final state is not necessarily
the ground state of the system $(^3\mathrm{H},\hspace{0mm}^3\mathrm{He}^+)$.
According to~\cite{tritium-sensitivity} the effect on the spectrum is
very large---larger than $10\,\%$ close to the endpoint. Far from the endpoint
($E_e^\mathrm{max}-E_e>1\,\text{keV}$) the corrections are estimated
to still be of the order of $1\,\%$, but expected to be smooth in $E_e$, since the
excitation energies of the $(^3\mathrm{H},\hspace{0mm}^3\mathrm{He}^+)$-system
are all below $200\,\text{eV}<1\,\text{keV}$~\cite{tritium-sensitivity}.
 \item Coulomb interaction between the outgoing electron, the daughter nucleus
($\rightarrow$ Fermi function $F(Z,\,E_e)$)
and the left behind orbital electron of the former $^3\mathrm{H}_2$-molecule.
 \item The nuclear recoil: This effect is automatically taken into account by
using the exact relativistic expression~(\ref{relativistic_spectrum}) for $d\Gamma/dE_e$.
 \item The daughter nucleus $^3\mathrm{He}^{2+}$ is not pointlike, which modifies
the Coulomb field acting on the emitted electron.
 \item Radiative corrections: The dominant radiative corrections will be QED-corrections
of the order of $\sim 1\,\%$.
\end{itemize}
All these corrections to the electron spectrum are estimated in~\cite{tritium-sensitivity}
and can be (at least far from the endpoint) assumed to be smooth. Moreover, ref.~\cite{tritium-sensitivity}
provides a sensitivity study showing that the maximal sensitivity of a KATRIN-like experiment
to the existence of keV sterile neutrinos will be diminished by these theoretical
uncertainties by a factor of only about 5 from the purely statistical sensitivity.

Another type of Standard Model physics which has to be taken into account
are corrections from nuclear structure. Up to now we have mostly treated the involved
nuclei as pointlike and only (electro)weakly interacting. In reality, the nuclei are bound states
of nucleons which themselves are bound states of quarks. Effects from QCD are therefore
not negligible for high-precision studies. We can take these effects into
account via so-called hadronic matrix elements.

\paragraph{Hadronic matrix elements:} At the quark level
the weak interaction Lagrangian for beta decay is given by
\begin{equation}
-\frac{G_F}{\sqrt{2}} V_{ud}
\left( \overline{e} \gamma^\mu (\mathbbm{1}-\gamma^5) \nu_e \right)
\left( \overline{u} \gamma_\mu (\mathbbm{1}-\gamma^5) d \right) + \mathrm{H.c.}
\end{equation}
In order to take into account that the initial and final states do not involve
single quarks but hadrons, instead of
$\langle u(p_u)| \left( \overline{u} \gamma_\mu (\mathbbm{1}-\gamma^5) d \right) |d(p_d)\rangle$
one has to consider the hadronic matrix element
\begin{equation}\label{def_hadronic}
\langle \mathcal{B}(p_\mathcal{B})|
\left( \overline{u} \gamma_\mu (\mathbbm{1}-\gamma^5) d \right)
|\mathcal{A}(p_\mathcal{A})\rangle,
\end{equation}
where $|\mathcal{A}(p_\mathcal{A})\rangle$ and $|\mathcal{B}(p_\mathcal{B})\rangle$
are the initial and final hadronic state, respectively.
Hadronic matrix elements 
are calculated by matching the low-energy effective theory of QCD
to the quark-level Lagrangian.\footnote{See~\cite{cirigliano,Cirigliano:2012ab} for a
more detailed discussion and a collection
of references.} Note that in section~\ref{new_physics} we will
use a quark-level Lagrangian containing also terms apart from the simple $V-A$ term
$\gamma^\mu(\mathbbm{1}-\gamma^5)$ and therefore 
we will also need the hadronic matrix elements for these terms.

Reference~\cite{cirigliano}
gives all relevant hadronic matrix elements for neutron beta decay and discusses their
relevance by ordering the individual contributions in terms
of powers of $q/M_N$, where
\begin{equation}
 q \equiv p_n-p_p,
 \quad
 M_N \equiv (m_n+m_p)/2.
\end{equation}
For tritium decay we have $q < 20\,\text{keV}$ and $M_N$ has to be replaced by
$M_N \equiv (\mA+\mB)/2 \simeq 3\,\text{GeV}$.
Thus we have $q/M_N < 10^{-5}$.
Since the sensitivity of a future KATRIN-like experiment to $d\Gamma/dE_e$ will
not be higher than $10^{-8}$~\cite{tritium-sensitivity}, we only need to take into account contributions
up to order $q/M_N$ and this also only for the Standard Model $V-A$
interaction.\footnote{All new physics interactions will have small
coupling constants which further suppress $q/M_N < 10^{-5}$.}
The relevant matrix elements are then given by~\cite{cirigliano}:
\begin{subequations}\label{hadronic}
\begin{align}
& \langle p(p_p)| \overline{u}\gamma_\mu d |n(p_n) \rangle =
\overline{u}_p(p_p) \left[ g_V(q^2)\gamma_\mu -i\frac{g_\mathrm{WM}(q^2)}{2M_N}\sigma_{\mu\nu}q^\nu
 \right] u_n(p_n) + \mathcal{O}((q/M_N)^2),\\
& \langle p(p_p)| \overline{u}\gamma_\mu\gamma_5 d |n(p_n) \rangle =
g_A(q^2)\,\overline{u}_p(p_p) \gamma_\mu \gamma_5 u_n(p_n) + \mathcal{O}((q/M_N)^2),\\
& \langle p(p_p)| \overline{u}d |n(p_n) \rangle =
g_S(q^2)\, \overline{u}_p(p_p) u_n(p_n),\\
& \langle p(p_p)| \overline{u} \gamma_5 d |n(p_n) \rangle = g_P(q^2)\, \overline{u}_p(p_p) \gamma_5 u_n(p_n) = \mathcal{O}(q/M_N),\\
& \langle p(p_p)| \overline{u} \sigma_{\mu\nu} d |n(p_n) \rangle = g_T(q^2)\, \overline{u}_p(p_p) \sigma_{\mu\nu} u_n(p_n)+ \mathcal{O}(q/M_N).
\end{align}
\end{subequations}
Thus, apart from multiplication of the individual quark-interactions with form
factors $g_V$, $g_A$, $g_S$, $g_P$ and
$g_T$,\footnote{Also the pseudotensor contribution will
occur in the Lagrangian of interest in section~\ref{new_physics}.
Using the identity
$\sigma^{\mu\nu}\gamma^5 = \frac{i}{2} \varepsilon^{\mu\nu\rho\sigma} \sigma_{\rho\sigma}$,
we find that the pseudotensor contribution obtains the same form factor $g_T$ as the
tensor contribution.}
the only relevant new term is
the weak magnetism contribution
\begin{equation}
\langle p(p_p)| \overline{u}\gamma_\mu d |n(p_n) \rangle_\text{WM} =
-i\frac{g_\mathrm{WM}(q^2)}{2M_N}
\overline{u}_p(p_p) \sigma_{\mu\nu}q^\nu u_n(p_n).
\end{equation}
In our framework based on the hadron model of~\cite{Simkovic},
we use equations~(\ref{hadronic}) with $n$ replaced by
$^3\mathrm{H}^+$ and $p$ replaced by $^3\mathrm{He}^{2+}$.
Note that the form factors are dependent on $q^2$. However,
in the first approximation this dependence will be of the form~\cite{Simkovic}
\begin{equation}
g_X(q^2) = \frac{g_X(0)}{\left(1-\frac{q^2}{M_X^2}\right)^2},
\end{equation}
where $M_X\sim 1\,\mathrm{GeV}$ is a cutoff scale ($X=V,\,A,\,S,\,P,\,T$). Therefore,
\begin{equation}
g_X(q^2) = g_X(0) \left( 1 + 2 \frac{q^2}{M_X^2} + \mathcal{O}((q/M_X)^4)\right).
\end{equation}
For tritium decay we have $q<20\,\mathrm{keV}$ and thus $2 \frac{q^2}{M_X^2} \lesssim 10^{-9}$,
\textit{i.e.}\ we can safely ignore the $q^2$-dependence of the form factors.

From the discussion here we see that an exact treatment of $\beta$-decay involving hadrons 
requires the weak magnetism term involving the tensor coupling $g_\mathrm{WM}$. 
Instead of giving the lengthy full matrix element, we simply write down the 
corrections to the parameters $A$,$\ldots$,$D_2$ of equation~(\ref{SM_para}). 
Neglecting all terms suppressed by $q^2/M_N^2$ we obtain:
\begin{subequations}
\begin{align}
\begin{split}
\Delta A / \gamma & = -\frac{g_\mathrm{WM} g_V \mA}{M_N}
\big(2 \mA^4+\mA^2 \left(-4 \mB^2+m_e^2+m_j^2\right) + \\
& \hspace{31mm} 2 \mB^4-\mB^2 \left(m_e^2+m_j^2\right)-\left(m_e^2-m_j^2\right)^2\big),
\end{split} \\
\begin{split}
\Delta B_1 / \gamma & = \frac{g_\mathrm{WM} \mA}{2 M_N}
\big(g_V \left(3 \mA^3+\mA^2 \mB+\mA \left(m_j^2-3
   \mB^2\right)-\mB^3+\mB m_e^2\right) \\
& \hspace{22mm}-g_A (\mA+\mB)
   \left(\mA^2-\mB^2+m_e^2-m_j^2\right)\big),
\end{split} \\
\begin{split}
\Delta B_2/\gamma & = \frac{g_\mathrm{WM} \mA}{2 M_N} \big(g_A (\mA+\mB) \left(\mA^2-\mB^2-m_e^2+m_j^2\right)+ \\
& \hspace{23.5mm} g_V \left(3 \mA^3+\mA^2 \mB+\mA \left(m_e^2-3 \mB^2\right)-\mB^3+\mB
   m_j^2\right)\big),
\end{split} \\
\Delta C/\gamma & = -\frac{2 g_\mathrm{WM} g_V \mA^2 (\mA+\mB)}{M_N},\label{DeltaC_analytic} \\
\Delta D_1/\gamma & = \frac{g_\mathrm{WM} \mA^2 (g_A-g_V) (\mA+\mB)}{M_N}, \\
\Delta D_2/\gamma & = -\frac{g_\mathrm{WM} \mA^2 (g_A+g_V) (\mA+\mB)}{M_N},
\end{align}
\end{subequations}
where $\gamma=16\, G_F^2\, |V_{ud}|^2\, |U_{ej}|^2$. 
We see in particular that $C$ is no longer zero. 

%% file: NP.tex
\section{\label{sec:NP}Contributions of new physics}\label{new_physics}

Having summarized the general kinematic structure and the properties of 
$\beta$-spectra in tritium, we can finally study the effect of possible beyond the 
Standard Model charged current contributions. 

\subsection{Effective operator approach}

In parameterizing new physics contributions to the 
beta decay amplitude~\cite{Lee-Yang,Jackson}, we use the
standard expansion of generic $4\times 4$-matrices in terms of the sixteen operators
\begin{subequations}\label{eq:O}
\begin{align}
& L \equiv \bone - \gamma^5,\quad
R \equiv \bone + \gamma^5,\\
& L_\mu \equiv \gamma_\mu L,\quad
R_\mu \equiv \gamma_\mu R,\\
& L_{\mu\nu} \equiv \sigma_{\mu\nu} L,\quad
R_{\mu\nu} \equiv \sigma_{\mu\nu} R,
\end{align}
\end{subequations}
where $\sigma_{\mu\nu} = \frac{i}{2}[\gamma_\mu,\gamma_\nu]$.
We use the notation of~\cite{cirigliano,Cirigliano:2012ab} to parameterize
possible new physics contributions to the charged-current interactions
at the level of dimension-six operators:\footnote{We are not
considering extremely exotic new physics such as violation of CPT or
Lorentz invariance~\cite{Diaz:2013saa}.}
\begin{equation}\label{Lagrangian}
\mathcal{L}_\mathrm{CC} = -\frac{G_F V_{ud}}{\sqrt{2}}
\left\{
 (1+\delta_\beta)(\overline{e} L_\mu \nu_e)(\overline{u} L^\mu d) +
 \sum_{j} \stackrel{(\sim)}{\epsilon_j} (\overline{e} \, \mathcal{O}_j \, \nu_e) (\overline{u} \, \mathcal{O}_j' \, d)
\right\} + \hc
\end{equation}
with the $\stackrel{(\sim)}{\epsilon_j}$, $\mathcal{O}_j$ and $\mathcal{O}_j'$ given in 
table~\ref{newphysoperators}. Note that $\epsilon_L$ is 
equivalent to a total rescaling of the rate. 
\begin{table}
\begin{center}
\begin{tabular}{|ccc|}
\hline
$\stackrel{(\sim)}{\epsilon_j}$ & $\mathcal{O}_j$ & $\mathcal{O}'_j$\\
\hline
$\epsilon_L$ & $\gamma_\mu(\bone-\gamma_5)$ & $\gamma^\mu(\bone-\gamma_5)$\\
$\widetilde{\epsilon}_L$ & $\gamma_\mu(\bone+\gamma_5)$ & $\gamma^\mu(\bone-\gamma_5)$\\
$\epsilon_R$ & $\gamma_\mu(\bone-\gamma_5)$ & $\gamma^\mu(\bone+\gamma_5)$\\
$\widetilde{\epsilon}_R$ & $\gamma_\mu(\bone+\gamma_5)$ & $\gamma^\mu(\bone+\gamma_5)$\\
$\epsilon_S$ & $\bone-\gamma_5$ & $\bone$\\
$\widetilde{\epsilon}_S$ & $\bone+\gamma_5$ & $\bone$\\
$-\epsilon_P$ & $\bone-\gamma_5$ & $\gamma^5$\\
$-\widetilde{\epsilon}_P$ & $\bone+\gamma_5$ & $\gamma^5$\\
$\epsilon_T$ & $\sigma_{\mu\nu}(\bone-\gamma_5)$ & $\sigma^{\mu\nu}(\bone-\gamma_5)$\\
$\widetilde{\epsilon}_T$ & $\sigma_{\mu\nu}(\bone+\gamma_5)$ & $\sigma^{\mu\nu}(\bone+\gamma_5)$\\
\hline
\end{tabular}
\end{center}
\caption{Coupling constants and operators for the new physics contributions
to $\mathcal{L}_\textit{\rm CC}$ of the 
form $\stackrel{(\sim)}{\epsilon_j} (\overline{e} \, \mathcal{O}_j \, \nu_e) (\overline{u} \, \mathcal{O}_j' \, d)$.}\label{newphysoperators}
\end{table}
The fields $e$, $d$ and $u$ are the electron, down and up quark
mass eigenfields respectively. The field $\nu_{e}$ is the
electron neutrino flavour field, containing in principle admixtures of 
sterile states, see equation (\ref{eq:2}).
The first term $(\overline{e} L_\mu \nu_e)(\overline{u} L^\mu d)$
is the usual Standard Model contribution,
which is then multiplied by a factor $1+\delta_\beta$ taking into
account the non-QED
electroweak 
radiative corrections; 
$G_F = g^2/(4\sqrt{2}M_W^2)$ is the tree-level
Standard Model Fermi constant.

The general parameterization in equation (\ref{Lagrangian}) has been obtained by 
using all possible and relevant dimension-6 operators including Standard Model fields and 
right-handed neutrinos \cite{Cirigliano:2012ab}. In case no right-handed neutrinos, \textit{i.e.}\ Standard Model singlet fermions or sterile neutrinos, are present, 
the $\widetilde{\epsilon}_j$ are absent. The various dimension-6 operators could be generated 
by integrating out 
heavy particles
in renormalizable theories beyond the SM.\footnote{\label{fn:lr}For instance, within left-right symmetric 
theories~\cite{LR1,LR2,LR3,LR4,LR5}, one could write at leading order  
$\epsilon_L = 0$, $\epsilon_R = \widetilde{\epsilon}_L = -\xi e^{-i\alpha}$ and  
$\widetilde{\epsilon}_R = M_{W_1}^2 /M_{W_2}^2$. Here $\xi$ and $\alpha$ appear as 
parameters linking the vector bosons of $SU(2)_L$ and $SU(2)_R$ with their mass eigenstates 
\begin{equation}\nonumber 
\begin{pmatrix}
W_L^\pm \\ W_R^\pm
\end{pmatrix} =
\begin{pmatrix}
\cos\xi & \sin\xi \, e^{i\alpha} \\
-\sin\xi \, e^{-i\alpha} & \cos\xi
\end{pmatrix}
\begin{pmatrix}
W_1^\pm \\ W_2^\pm
\end{pmatrix}.
\end{equation}
}

We see that there are in principle 
ten 
additional charged-current 
contributions to $\beta$-decay, two of which ($\epsilon_L$ and $\widetilde{\epsilon}_R$)
have the same Lorentz structure 
as the Standard Model term, while the other eight enjoy a non-SM structure. We will analyze the 
effect of the ten operators on the electron energy spectrum for both light ($\ls 0.5$ eV) and 
heavy ($\ls 10$ keV) neutrinos.

\subsection{Neutrino mass and flavour eigenstates}\label{numass_flavour}

In extensions of the Standard Model with
right-handed neutrinos, the terms of~(\ref{Lagrangian})
which contain right-handed neutrino fields in general do not vanish,
\textit{i.e.}\
\begin{equation}
\widetilde{\epsilon}_L,\,
\widetilde{\epsilon}_R,\,
\widetilde{\epsilon}_S,\,
\widetilde{\epsilon}_P,\,
\widetilde{\epsilon}_{T}\neq 0.
\end{equation}
However, $\epsilon_P$ and $\widetilde{\epsilon}_P$
come along with the pseudoscalar contribution to the hadronic
matrix element (see equation~(\ref{hadronic}d)) which is suppressed by
a factor of $q/M_N\sim 10^{-5}$. Even though the pseudoscalar coupling
$g_P$ is rather large (see table~\ref{masses}), the suppression is not compensated and 
effects of $\epsilon_P$ and $\widetilde{\epsilon}_P$ are negligible for
heavy neutrinos and quite small for light neutrinos---see section~\ref{numerical_analysis}.

Generically allowing all three types of neutrino mass terms (Dirac, type-I seesaw, type-II seesaw),
the neutrino mass term is given by
\begin{equation}\label{nu-massterm}
\mathcal{L}_{\nu}^\text{mass} = -\frac{1}{2}\overline{n_L} M_\nu n_L^c +\hc,
\end{equation}
where
\begin{equation}
\label{massmat}
M_\nu =
\begin{pmatrix}
M_L & M_D \\
M_D^T & M_R
\end{pmatrix}
\mbox{ and } 
n_L=
\begin{pmatrix}
\nu_L\\ \nu_R^c
\end{pmatrix}.
\end{equation}
We use the notation of~\cite{Barry}, where $M_\nu$ is diagonalized via
\begin{equation}
W^\dagger M_\nu W^\ast = \mathrm{diag}(m_1,\,m_2,\,m_3,\,M_1,\,M_2,\,M_3).
\end{equation}
Here
\begin{equation}\label{STUV}
W = \begin{pmatrix}
U & S\\
T & V
\end{pmatrix}
\end{equation}
is unitary and $m_i$ are the masses of the light neutrino mass eigenfields $\nu_i'$
and $M_j$ are the masses of the heavy neutrino mass eigenfields $N_{Rj}'$.
The neutrino flavour fields are then given by
\begin{subequations}\label{light-heavy}
\begin{align}
& \nu_L = U \nu_L' + S N_R'\hspace{0mm}^c,\\
& \nu_R = T^\ast \nu_L'\hspace{0mm}^c + V^\ast N_R'.
\end{align}
\end{subequations}
Since for massive Majorana fields we have
$\nu^c=\nu$ and $N^c = N$,
equation~(\ref{light-heavy}) simplifies to
\begin{subequations}
\begin{align}
& \nu_L = U \nu_L' + S N_L',\\
& \nu_R = T^\ast \nu_R' + V^\ast N_R'.
\end{align}
\end{subequations}
The ``left-right mixing'' matrices $S$ and $T$ are constrained to be small and will 
suppress interactions of right-handed neutrinos. Nevertheless, interesting 
effects can arise.

\subsection{The energy spectrum}

The Lagrangian~(\ref{Lagrangian}) has eleven individual terms (the Standard Model term and ten new
physics contributions). In the following we evaluate the expressions for the individual contributions
to the matrix element of $\beta$-decay involving heavy, \textit{i.e.}\ few keV, 
and light antineutrinos in the final state, respectively.
The amplitude $\mathcal{M}$ for the decay is given by
\begin{equation}
\mathcal{M} = -\frac{G_F V_{ud}}{\sqrt{2}}
\sum_\alpha
C^{(\alpha)}\, X_{ej}^{(\alpha)}\, [\overline{u}_e \mathcal{O}^{(\alpha)} v_j]
[\overline{u}_\mathcal{B} \mathcal{O}^{(\alpha)}\hspace{0mm}' u_\mathcal{A}],
\end{equation}
where $C^{(\alpha)}$ is a constant, $X_{ej}^{(\alpha)}$ is the $ej$-element of the 
(in general non-unitary) mixing matrix $X^{(\alpha)}$ ($X^{(\alpha)}=U,S,T^\ast,V^\ast$)
and $\mathcal{O}^{(\alpha)}$ and $\mathcal{O}^{(\alpha)}\hspace{0mm}'$ are $4\times 4$-matrices, see 
equation (\ref{eq:O}) and table \ref{newphysoperators}.
The index $\alpha$ runs over the eleven contributions to $\mathcal{L}_\mathrm{CC}$ of
equation~(\ref{Lagrangian}).
The contributions to the matrix element for the emission of a light
antineutrino mass eigenstate $|\overline{\nu}_j\rangle$ or heavy
antineutrino mass eigenstate $|\overline{N}_j\rangle$ are shown
in table~\ref{M_light_nu}.

\begin{table}
\begin{center}
\begin{tabular}{|l||l|l|l|l|}
\hline
 $\alpha$ & $C^{(\alpha)}$ & $X_{ej}^{(\alpha)}$ & $\mathcal{O}^{(\alpha)}$ & $\mathcal{O}^{(\alpha)}\hspace{0mm}'$\\
\hline
\hline
 SM & $(1+\delta_\beta)$ & $U_{ej}$ & $\gamma_\mu (\mathbbm{1}-\gamma^5)$ 
& $g_V\gamma^\mu - i\frac{g_\mathrm{WM}}{2M_N} \sigma^{\mu\nu} q_\nu -g_A \gamma^\mu \gamma^5$\\
\hline
 $\epsilon_L$ & $\epsilon_L$ & $U_{ej}$ & $\gamma_\mu (\mathbbm{1}-\gamma^5)$
& $g_V\gamma^\mu - i\frac{g_\mathrm{WM}}{2M_N} \sigma^{\mu\nu} q_\nu -g_A \gamma^\mu \gamma^5$\\
\hline
 $\widetilde{\epsilon}_L$ & $\widetilde{\epsilon}_L$ & $T_{ej}^\ast$ & $\gamma_\mu (\mathbbm{1}+\gamma^5)$ 
& $g_V\gamma^\mu - i\frac{g_\mathrm{WM}}{2M_N} \sigma^{\mu\nu} q_\nu -g_A \gamma^\mu \gamma^5$\\
\hline
 $\epsilon_R$ & $\epsilon_R$ & $U_{ej}$ & $\gamma_\mu (\mathbbm{1}-\gamma^5)$ 
& $g_V\gamma^\mu - i\frac{g_\mathrm{WM}}{2M_N} \sigma^{\mu\nu} q_\nu +g_A \gamma^\mu \gamma^5$\\
\hline
 $\widetilde{\epsilon}_R$ & $\widetilde{\epsilon}_R$ & $T_{ej}^\ast$ & $\gamma_\mu (\mathbbm{1}+\gamma^5)$ 
& $g_V\gamma^\mu - i\frac{g_\mathrm{WM}}{2M_N} \sigma^{\mu\nu} q_\nu +g_A \gamma^\mu \gamma^5$\\
\hline
 $\epsilon_S$ & $\epsilon_S$ & $U_{ej}$ & $\mathbbm{1}-\gamma^5$ 
& $g_S \mathbbm{1}$\\
\hline
 $\widetilde{\epsilon}_S$ & $\widetilde{\epsilon}_S$ & $T_{ej}^\ast$ & $\mathbbm{1}+\gamma^5$ 
& $g_S \mathbbm{1}$\\
\hline
 $\epsilon_P$ & $-\epsilon_P$ & $U_{ej}$ & $\mathbbm{1}-\gamma^5$
& $g_P \gamma^5$\\
\hline
 $\widetilde{\epsilon}_P$ & $-\widetilde{\epsilon}_P$ & $T_{ej}^\ast$ & $\mathbbm{1}+\gamma^5$ 
& $g_P \gamma^5$\\
\hline
 $\epsilon_T$ & $\epsilon_T$ & $U_{ej}$ & $\sigma_{\mu\nu}(\mathbbm{1}-\gamma^5)$ 
& $g_T \sigma^{\mu\nu} (\mathbbm{1}-\gamma^5)$\\
\hline
 $\widetilde{\epsilon}_T$ & $\widetilde{\epsilon}_T$ & $T_{ej}^\ast$ & $\sigma_{\mu\nu}(\mathbbm{1}+\gamma^5)$
& $g_T \sigma^{\mu\nu} (\mathbbm{1}+\gamma^5)$\\
\hline
\end{tabular}
\end{center}
\caption{The contributions to the matrix element
$\mathcal{M}(\mathcal{A}\rightarrow \mathcal{B} + e^- + \overline{\nu}_j)$. Here 
$q=p_\mathcal{A}-p_\mathcal{B}$ and $M_N=(\mA+\mB)/2$. The expression for
$\mathcal{M}(\mathcal{A}\rightarrow \mathcal{B} + e^- + \overline{N}_j)$ is the
same with the replacements $U\rightarrow S$ and $T\rightarrow V$.}\label{M_light_nu}
\end{table}

The amplitude squared averaged over the spin orientation of the decaying nucleus
and summed over the spins of all final state particles is then given by
\begin{equation}
    |\mathcal{M}|^2 = \frac{1}{2}\sum_\mathrm{spins} |\mathcal{M}(\mathrm{spins})|^2 =
    \sum_{\alpha,\beta} S_{\alpha\beta}\,,
\end{equation}
where
\begin{equation}
\begin{split}
  S_{\alpha\beta} \equiv \, & \frac{G_F^2 |V_{ud}|^2}{4}
  C^{(\alpha)} C^{(\beta)\ast} X_{ej}^{(\alpha)} X_{ej}^{(\beta)\ast} \times \\
 & \mathrm{Tr} \left[ (\slashed{p}_e + m_e) \mathcal{O}^{(\alpha)} (\slashed{p}_j-m_j) (\gamma^0 \mathcal{O}^{(\beta)\dagger} \gamma^0 )\right] \times\\
 & \mathrm{Tr} \left[ (\slashed{p}_\mathcal{B} + \mB) \mathcal{O}^{(\alpha)}\hspace{0mm}'
(\slashed{p}_\mathcal{A}-\mA) (\gamma^0 \mathcal{O}^{(\beta)}\hspace{0mm}'\hspace{0mm}^\dagger \gamma^0 )\right].
\end{split}\label{S}
\end{equation}
Defining that $\alpha < \beta$ if $\alpha$ comes before $\beta$ in table~\ref{M_light_nu}, we
can rewrite this as a sum with purely real summands:
\begin{equation}
 |\mathcal{M}|^2 = \sum_{\alpha} S_{\alpha\alpha} + 2 \sum_{\alpha < \beta} \mathrm{Re}\, S_{\alpha\beta}.
\end{equation}
Thus the total matrix element squared consists of
66 real terms.
For each of these 66 terms we can take the traces. Removing all terms suppressed by
factors $(q/M_N)^2$, the terms of the form~(\ref{sixmomenta}) are removed automatically.
Using energy momentum conservation we can then compute the six parameters $A$ to $D_2$ of equation~(\ref{parameterization}).
Since the 66 summands are real, also $A$ to $D_2$ are real for each summand.
For the trace computations we used \textit{Package-\textbf{X}}~\cite{X1,X2}.
Note that the product of the two traces in equation~(\ref{S}) is real. Hence,
\begin{equation}
  \mathrm{Re}\,S_{\alpha\beta} \propto \mathrm{Re} \left(C^{(\alpha)} C^{(\beta)\ast} X_{ej}^{(\alpha)} X_{ej}^{(\beta)\ast}\right).
\end{equation}
All matrix elements have the general parameterization in terms of 
$A,B_{1,2},C,D_{1,2}$ from equation (\ref{parameterization}), but of course they are 
different from their Standard Model expressions. The parameters $A,B_{1,2},C,D_{1,2}$ 
possess three contributions, the Standard Model term (SM), 
the New Physics term (NP$^2$) and 
their interference term (SM-NP).
A general property worth mentioning is that the interference terms of the $\widetilde{\epsilon}$ 
operators vanish for $m_j\rightarrow 0$, which can easily be understood from
chirality considerations. 
We will not give the lengthy full expressions for all terms, let us 
simply give two illustrative examples. Defining the overall constant
\begin{equation}
\gamma' = 2 \, G_F^2 \, |V_{ud}|^2 \, |U_{ej}|^2 \, \mathrm{Re}\left( (1+\delta_\beta) \epsilon_R^\ast \right),
\end{equation}
we obtain for the coefficients for the SM-$\epsilon_R$ interference term:
\begin{subequations}
\begin{align}
\begin{split}
A/\gamma' = \; & 8 \mA \mB \left(g_A^2+g_V^2\right) \left(\mA^2-\mB^2+m_e^2+m_j^2\right) \\
& -\frac{4 g_\mathrm{WM} g_V \mA}{M_N} (2 \mA^4+\mA^2 \left(-4 \mB^2+m_e^2+m_j^2\right)+ \\
& \hspace{29.5mm} 2\mB^4-\mB^2 \left(m_e^2+m_j^2\right)-\left(m_e^2-m_j^2\right)^2),
\end{split} \\
\begin{split}
B_1/\gamma' = \; & -8 \mA (g_A^2 \left(\mA^2+2 \mA \mB-\mB^2+m_e^2-m_j^2\right)+ \\
& \hspace{15mm} g_V^2 \left(-\mA^2+2 \mA \mB+\mB^2-m_e^2+m_j^2\right)) \\
& + \frac{8 g_\mathrm{WM} g_V \mA}{M_N} \left(3 \mA^3+\mA^2 \mB+\mA \left(m_j^2-3 \mB^2\right)-\mB^3+\mB m_e^2\right),
\end{split} \\
B_2/\gamma' = \; & B_1/\gamma' \vert_{m_e \leftrightarrow m_j}, \\
C/\gamma' = \; & -\frac{32 g_\mathrm{WM} g_V \mA^2 (\mA+\mB)}{M_N}, \\
D_1/\gamma' = \; & 16 \mA^2 (g_A-g_V) (g_A+g_V)
   -\frac{16 g_\mathrm{WM} g_V \mA^2 (\mA+\mB)}{M_N}, \\
D_2/\gamma' = \; &  D_1/\gamma'.
\end{align}
\end{subequations}
For the SM-$\widetilde{\epsilon}_R$ interference contribution 
(corresponding to right-handed currents, see footnote \ref{fn:lr}) one obtains:
\begin{equation}
\gamma'' = 2 \, G_F^2 \, |V_{ud}|^2 \, \mathrm{Re}\left( U_{ej} T_{ej} (1+\delta_\beta) \widetilde{\epsilon}_R^\ast \right),
\end{equation}
\begin{subequations}
\begin{align}
\begin{split}
A/\gamma'' = \; & 16 \mA m_e m_j \left(g_A^2 (-\mA)-2 g_A^2 \mB+g_V^2 \mA-2 g_V^2 \mB\right) \\
& +\frac{24 g_\mathrm{WM} g_V \mA m_e
   m_j (\mA-\mB) (\mA+\mB)}{M_N},
\end{split} \\
B_1/\gamma'' = \; & -16 \mA m_e m_j (g_V-g_A) (g_A+g_V)-\frac{24 g_\mathrm{WM} g_V \mA m_e m_j (\mA+\mB)}{M_N}, \\
B_2/\gamma'' = \; & B_1/\gamma'', \\
C/\gamma'' = \; & D_1/\gamma'' = D_2/\gamma'' = 0.
\end{align}
\end{subequations}
Due to the
different chiralities 
of the neutrino fields (left in the SM term and right in the
new physics contribution $\propto\widetilde{\epsilon}_R$), as expected, all coefficients are
proportional to the neutrino mass $m_j$. Consequently, these interference terms are suppressed
for the emission of light neutrinos, but play an important role if heavy neutrinos are emitted.

Next we will perform a numerical study of the possible 
corrections to $A,B_{1,2},C,D_{1,2}$ with respect to their form in the Standard Model, 
and also plot the relative deviation from the shape of the Standard Model electron energy 
spectrum.

\subsection{Numerical analysis}\label{numerical_analysis}

The values for the masses, SM coupling constants and form factors
we use for the computation of the tritium beta spectrum are shown in
table~\ref{masses}.
Current bounds on the real and imaginary parts of the new-physics coupling
constants $\epsilon$ and $\widetilde{\epsilon}$ are given in~\cite{cirigliano,Cirigliano:2012ab}.
All constraints are compatible with zero values for these constants.
However, the bounds differ in their orders of magnitude (from $|\mathrm{Im}\,\epsilon_P| < 2\times 10^{-4}$
to $|\mathrm{Re}\,\widetilde{\epsilon}_L| < 6\times 10^{-2}$ at $90\,\%$ CL~\cite{cirigliano}).
The bounds we use are shown in table~\ref{epsilons}.
\begin{table}
\begin{center}
\renewcommand{\arraystretch}{1.2}
\begin{tabular}{|l|l|l|}
\hline
Quantity & value & comment/reference\\
\hline
mass of $e^{-}$ & 0.510998928(11) MeV & \cite{PDG} \\
mass of $^3$H (atom) & 2809.43185(11) MeV & \cite{AME2003-1,AME2003-2}\\
mass of $^3$H$^+$ (nucleus) & 2808.92085(11) MeV & \\
mass of $^3$He (atom) & 2809.41325(11) MeV & \cite{AME2003-1,AME2003-2}\\
mass of $^3$He$^{2+}$ (nucleus) & 2808.39126(11) MeV & \\
$G_F$ & $1.1663787(6) \times 10^{-5}\,\mathrm{GeV}^{-2}$ & \cite{PDG}\\
$g_V$ & $1.0$ & CVC hypothesis~\cite{CVC1,CVC2}\\
$g_A/g_V$ &  $1.2646 \pm 0.0035$ & \cite{akulov}\\
$|V_{ud}|$ & $0.97425 \pm 0.00022$ & \cite{PDG}\\
$g_S$ & $1.02 \pm 0.11$ & $\overline{\mathrm{MS}}, \mu=2\,\text{GeV~\cite{Gonzalez-Alonso:2013ura}}$\\
$g_P$ & $349 \pm 9$ & $\overline{\mathrm{MS}}, \mu=2\,\text{GeV~\cite{Gonzalez-Alonso:2013ura}}$\\
$g_T$ & $1.020 \pm 0.076$ & $\text{lattice, }\overline{\mathrm{MS}}, \mu=2\,\text{GeV~\cite{Bhattacharya:2015esa}}$\\
$g_\mathrm{WM}$ & $-6.106$ & \cite{Simkovic}\\
\hline
\end{tabular}
\renewcommand{\arraystretch}{1.0}
\caption{The quantities needed for the numerical computation
of the electron energy spectrum of tritium beta decay.
For the computation of the nuclei masses we have neglected the
binding energy of the electrons (which is $<100~\text{eV}$). The errors include the
error of the determination of the atomic mass
unit $u=(931494.013\pm0.037)~\text{keV}$~\cite{AME2003-1}.
(CVC = Conserved Vector Current).}\label{masses}
\end{center}
\end{table}
The six parameters for the Standard Model contribution $S_{\mathrm{SM},\mathrm{SM}}$ using the numerical
input from table~\ref{masses}, setting $\delta_\beta=0$ and assuming only three massless neutrino states with
$\sum_{j=1}^3 |U_{ej}|^2 = 1$ are
\begin{subequations}\label{SM-values}
\begin{align}
& A_{\mathrm{SM}} = -1.45 \times 10^{-11},\\
& B_{1,\mathrm{SM}} = 2.74 \times 10^{-11}\,\mathrm{MeV}^{-1},\\
& B_{2,\mathrm{SM}} = 2.71 \times 10^{-11}\,\mathrm{MeV}^{-1},\\
& C_{\mathrm{SM}} = 3.98 \times 10^{-13}\,\mathrm{MeV}^{-2},\\
& D_{1,\mathrm{SM}} = -5.38 \times 10^{-14}\,\mathrm{MeV}^{-2},\\
& D_{2,\mathrm{SM}} = 3.67 \times 10^{-13}\,\mathrm{MeV}^{-2}.
\end{align}
\end{subequations}
Note that the inclusion of the weak magnetism term proportional to $g_\mathrm{WM}$ in equation 
(\ref{hadronic}a) gives a small contribution to $C_{\rm SM}$, which is zero in the 
pure $V-A$ interaction case, cf.~(\ref{SM_para}) and~(\ref{DeltaC_analytic}).
In the following, when we compare parameters to the Standard Model values, we
\textit{always} mean the above numbers.

In order to get a feeling for the sizes of the physical effects,
we compute the relative sizes of the contributions to the total decay width, see table~\ref{decay_widths}.
As can be seen from this table, there are huge cancellations among the six different terms.
Therefore, in general all six contributions are important.

\begin{table}
\begin{center}
\begin{tabular}{|c|c|c|c|}
\hline
parameter & \multicolumn{2}{|c|}{best 90\,\% CL upper bound~\cite{cirigliano}} & used for our estimation\\
\hline
 & $|\mathrm{Re}\,\epsilon|$ & $|\mathrm{Im}\,\epsilon|$ & $\epsilon$ \\
\hline
\hline
$\epsilon_L$ & $5 \times 10^{-4}$ & $5 \times 10^{-3}$ & $5.0 \times 10^{-3}$ \\
$\widetilde{\epsilon}_L$ & $6 \times 10^{-2}$ & --- & $8.5 \times 10^{-2}$ \\
$\epsilon_R$ & $5 \times 10^{-4}$ & $5 \times 10^{-4}$ & $7.1 \times 10^{-4}$ \\
$\widetilde{\epsilon}_R$ & $5 \times 10^{-3}$ & $5 \times 10^{-3}$ & $7.1 \times 10^{-3}$ \\
$\epsilon_S$ & $8 \times 10^{-3}$ & $1 \times 10^{-2}$ & $1.3 \times 10^{-2}$ \\
$\widetilde{\epsilon}_S$ & $1.3 \times 10^{-2}$ & $1.3 \times 10^{-2}$ & $1.8 \times 10^{-2}$ \\
$\epsilon_P$ & $4 \times 10^{-4}$ & $2 \times 10^{-4}$ & $4.5 \times 10^{-4}$ \\
$\widetilde{\epsilon}_P$ & $2 \times 10^{-4}$ & $2 \times 10^{-4}$ & $2.8 \times 10^{-4}$ \\
$\epsilon_T$ & $1 \times 10^{-3}$ & $1 \times 10^{-3}$ & $1.4 \times 10^{-3}$ \\
$\widetilde{\epsilon}_T$ & $3 \times 10^{-3}$ & $3 \times 10^{-3}$ & $4.2 \times 10^{-3}$ \\
\hline
\end{tabular}
\end{center}
\caption{Numerical values for the coupling constants $\epsilon$ and $\widetilde{\epsilon}$
used for our analysis.}\label{epsilons}
\end{table}

\begin{table}
\begin{center}
\renewcommand{\arraystretch}{1.2}
\begin{tabular}{|l|r|}
\hline
Contribution from & $\Gamma/\Gamma_\mathrm{total}$\\
\hline
$A$ & $-26583.954$\\
$B_1$ & $26051.460$\\
$B_2$ & $555.755$\\
$C$ & $4.214$\\
$D_1$ & $-26.572$ \\
$D_2$ & $0.097$ \\
\hline
Sum & $1.000$\\
\hline
\end{tabular}
\renewcommand{\arraystretch}{1.0}
\caption{Contributions of the different terms in $|\mathcal{M}|^2$ in the Standard
Model to the total decay width of tritium. Note that since we have divided $|\mathcal{M}|^2$
into real (but not necessarily positive) parts, the different $\Gamma$ contributing
to $\Gamma_\mathrm{total}$ can have either sign.}\label{decay_widths}
\end{center}
\end{table}

In table~\ref{num1} 
(see appendix) 
we give the numerical values for the 
coefficients $A/A_{\rm SM}$ to $D_2/D_{2,\mathrm{SM}}$ for $U_{ej}=V_{ej}=S_{ej}=T_{ej}=1$
and $\epsilon=\widetilde{\epsilon}=1$. From these, the values of $A$ to $D_2$ can be computed
by multiplication with the Standard Model values of equation~(\ref{SM-values}) and the appropriate
suppression factors found in table~\ref{suppression}, to be discussed next.

\paragraph{Numerical values for the suppression factors:}
In order to estimate the suppression factors of table~\ref{suppression}, we use the 90\,\% CL
upper bounds on the $\epsilon$ and $\widetilde{\epsilon}$ from~\cite{cirigliano}. 
Since we only perform an order of magnitude estimation, we use the real positive values
$\epsilon = \sqrt{|\mathrm{Re}\,\epsilon|^2 + |\mathrm{Im}\,\epsilon|^2}$---see table~\ref{epsilons}.
In those cases where there is no bound on $|\mathrm{Im}\,\epsilon|$, we set $\epsilon = \sqrt{2} |\mathrm{Re}\,\epsilon|$.

Moreover, we have to fix at least the order of magnitude of the values of
the mixing matrix elements $U_{ej}$, $T_{ej}$,
$S_{ej}$ and $V_{ej}$. We use:
\begin{equation}
U_{ej} = V_{ej} = 1, \quad T_{ej} = S_{ej} = 10^{-3},
\end{equation}
which resembles a typical size of the effects of active-sterile mixing 
($i.e.$ proportional to eV/keV) compared to the ``active only'' values of
\begin{equation}
\left| \frac{S_{ej}}{U_{ej}} \right|^2 \sim 10^{-6}. 
\end{equation}
Values for other constraints can be easily obtained by
rescaling the results according to table~\ref{suppression}. 
Note that depending on the model, strong constraints on the mixing of keV-neutrinos may 
exist, for instance in the context of left-right symmetric theories and 
Warm Dark Matter decay, see for instance \cite{Barry}. We will however 
not go into detail here, but rather wish to show the shape of the spectral distortion of the 
new interactions and to demonstrate the capability of a modified 
KATRIN setup to give strong laboratory limits on 
exotic charged current interactions. 
We have by now all ingredients to make a full numerical comparison of the
sizes of the new-physics contributions to the electron energy spectrum.
Table~\ref{num2} 
(see appendix) 
shows, using the just discussed suppression factors, the sizes of the new physics effects for the emission
of a light ($m_j=0.5\,\text{eV}$) and heavy ($M_j=5\,\mathrm{keV}$) neutrino.

For illustration of the possible distortions of the electron 
spectra we first define reference spectra to which we will compare
different sample scenarios of new physics in beta decay.
\begin{itemize}
 \item \textbf{Reference spectrum 1N:} The beta decay spectrum
for only Standard Model interactions with three neutrinos and
normal mass hierarchy ($m_1=0$). For the mass-squared differences
and the values of the mixing angles
we use the values from the global fit~\cite{Schwetz}
which imply $m_\beta = 8.7\,\text{meV} \ll 0.2\,\text{eV}$.
Within KATRIN's experimental possibilities, this effectively corresponds to massless neutrinos.
 \item \textbf{Reference spectrum 1I:} The same spectrum as reference 1N, but with
an inverted mass hierarchy ($m_3=0$). For this case one obtains
$m_\beta = 48\,\text{meV} < 0.2\,\text{eV}$.
Within KATRIN's experimental possibilities, this is still not
distinguishable from the case of massless neutrinos. We note that Project 8 has 
in principle (using an atomic tritium source among other 
modifications) the option to reach such low values \cite{Doe:2013jfe}. 

\begin{table}[t]
\begin{center}
\begin{tabular}{|c|c|c|c|c|}
\hline
 & \multicolumn{2}{|c|}{light neutrino emission} & \multicolumn{2}{|c|}{heavy neutrino emission}\\
\hline
 & SM-NP & NP$^2$ & SM-NP & NP$^2$\\
$\epsilon$ & $\mathrm{Re}(\epsilon\,U_{ej}^2)$ & $|\epsilon|^2\,|U_{ej}|^2$ & $\mathrm{Re}(\epsilon\, S_{ej}^2)$ & $|\epsilon|^2\,|S_{ej}|^2$\\
$\widetilde{\epsilon}$ & $\mathrm{Re}(\widetilde{\epsilon}\,U_{ej} T_{ej}^\ast)$ & $|\widetilde{\epsilon}|^2\,|T_{ej}|^2$ & $\mathrm{Re}(\widetilde{\epsilon}\, S_{ej} V_{ej}^\ast)$ & $|\widetilde{\epsilon}|^2\,|V_{ej}|^2$\\
\hline
\end{tabular}
\end{center}
\caption{Suppression factors for the new-physics contributions to
the electron energy spectrum of beta decay.}\label{suppression}
\end{table}

 \item \textbf{Reference spectrum 2:} The beta decay spectrum
for only Standard Model interactions with one neutrino of 
mass $m_j=0.5\,\mathrm{eV}$ and $U_{ej} = 1$.
This corresponds to three quasi-degenerate light neutrinos with
$\sum_j |U_{ej}|^2=1$,
\textit{i.e.}\ a spectrum to be expected in KATRIN
if $m_\beta\approx 0.5\,\mathrm{eV} > 0.2\,\text{eV}$.
 \item \textbf{Reference spectrum 3:} Reference spectrum 2 (quasi-degenerate light neutrinos) plus 
one sterile neutrino with $M_j=5\,\mathrm{keV}$ and mixing $S_{ej}=10^{-3}$.
\end{itemize}

Our sample scenarios that include new physics are:
\begin{itemize}
 \item \textbf{Test spectrum AN:} The reference spectrum 1N (\textit{i.e.}\ left-handed neutrinos)
with new interactions $\epsilon_i \neq 0$. Since we neither add heavy nor right-handed neutrino fields,
there are no $\widetilde{\epsilon}$-interactions and $T_{ej}=S_{ej}=V_{ej}=0$.
 \item \textbf{Test spectrum AI:} Like the test spectrum AN but for an inverted neutrino mass hierarchy
($m_3=0$).
 \item \textbf{Test spectrum B:} Like the test spectra 1N and 1I, but this time for quasi-degenerate neutrinos
with $m_\beta \approx 0.5\,\text{eV}$.
 \item \textbf{Test spectrum C:} The reference spectrum 2 (three quasi-degenerate light neutrinos) but including 
a heavy neutrino ($M_j=5\,\mathrm{keV}$, $U_{ej}=T_{ej}=0$) having 
mixing matrix elements $S_{ej}=10^{-3}$, $V_{ej}=1$ and new interactions. 
\end{itemize}
We compare the test spectra to the reference spectra by plotting
\begin{equation}\label{eq:Del}
\Delta(\stackrel{(\sim)}{\epsilon_j})
\equiv \frac{\text{test spectrum} - \text{reference spectrum}}{\text{reference spectrum}} =
\frac{\text{test spectrum}}{\text{reference spectrum}}-1,
\end{equation}
where $\stackrel{(\sim)}{\epsilon_j}$ means that we turn on the
new physics parameter $\stackrel{(\sim)}{\epsilon_j}$
(setting all other epsilons to zero).  In all plots we set $\delta_\beta=0$.
The comparisons that were analyzed are:
\begin{itemize}
 \item Test spectrum AN (AI) \textit{vs.}\ reference spectrum 1N (1I). This shows the
effect of new physics in the worst case of extremely light neutrinos. 
 \item Test spectrum B \textit{vs.}\ reference spectrum 2. This shows the
effect of new physics for observable light neutrino emission, $m_\beta \approx 0.5$ eV 
in this case. 
 \item Test spectrum C \textit{vs.}\ reference spectrum 3. This shows the
effect of new physics for heavy neutrino emission, $M_j = 5$ keV and mixing 
$S_{ej} = 10^{-3}$ in this case. 

\end{itemize}

\begin{figure}
\begin{center}
\includegraphics[width=0.50\textwidth]{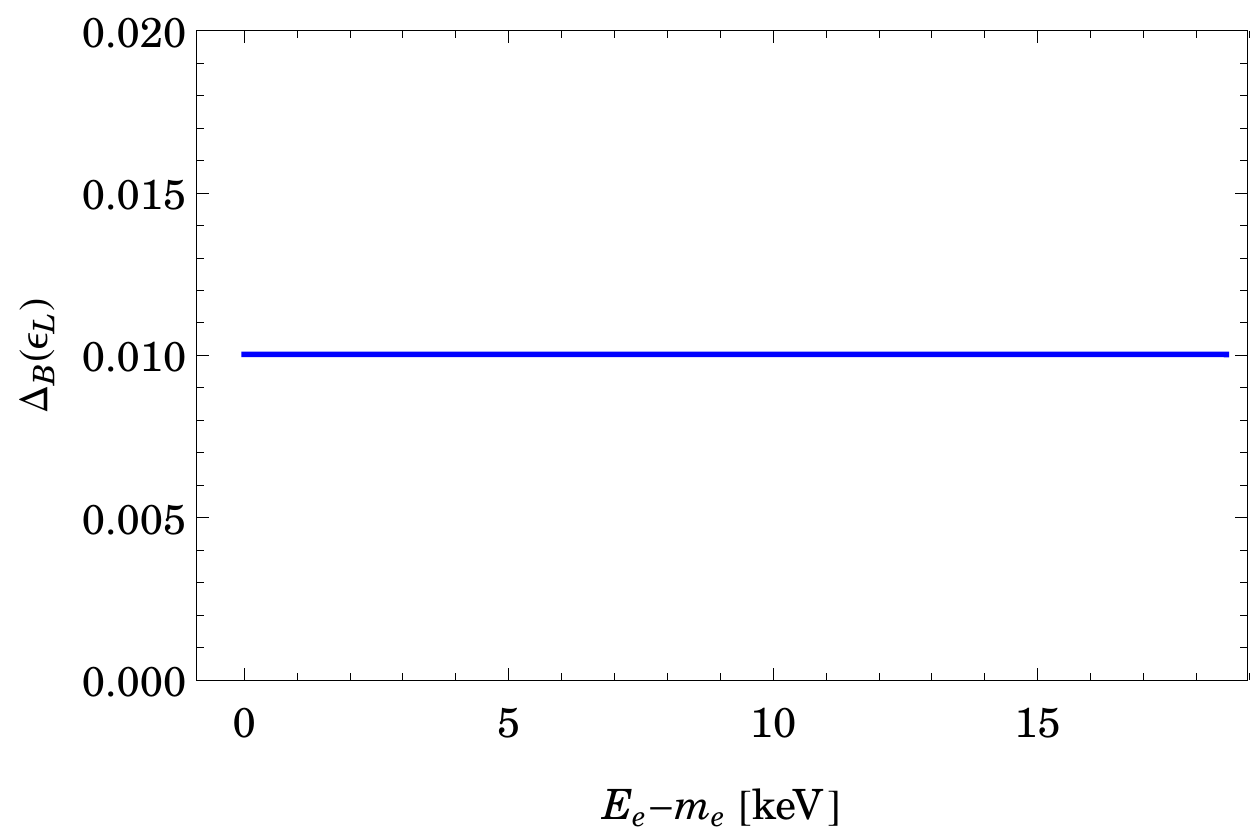}\hspace*{5mm}
\includegraphics[width=0.50\textwidth]{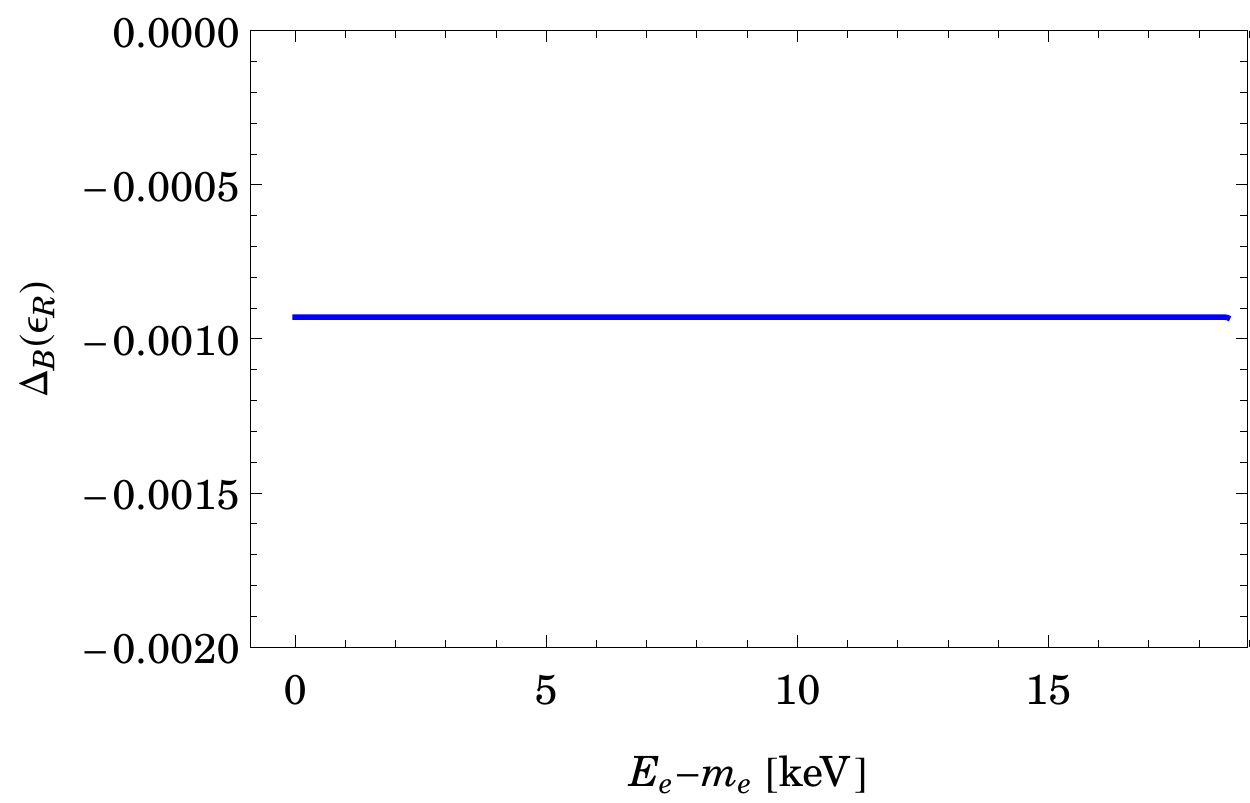}\\
\vspace*{3mm}
\includegraphics[width=0.50\textwidth]{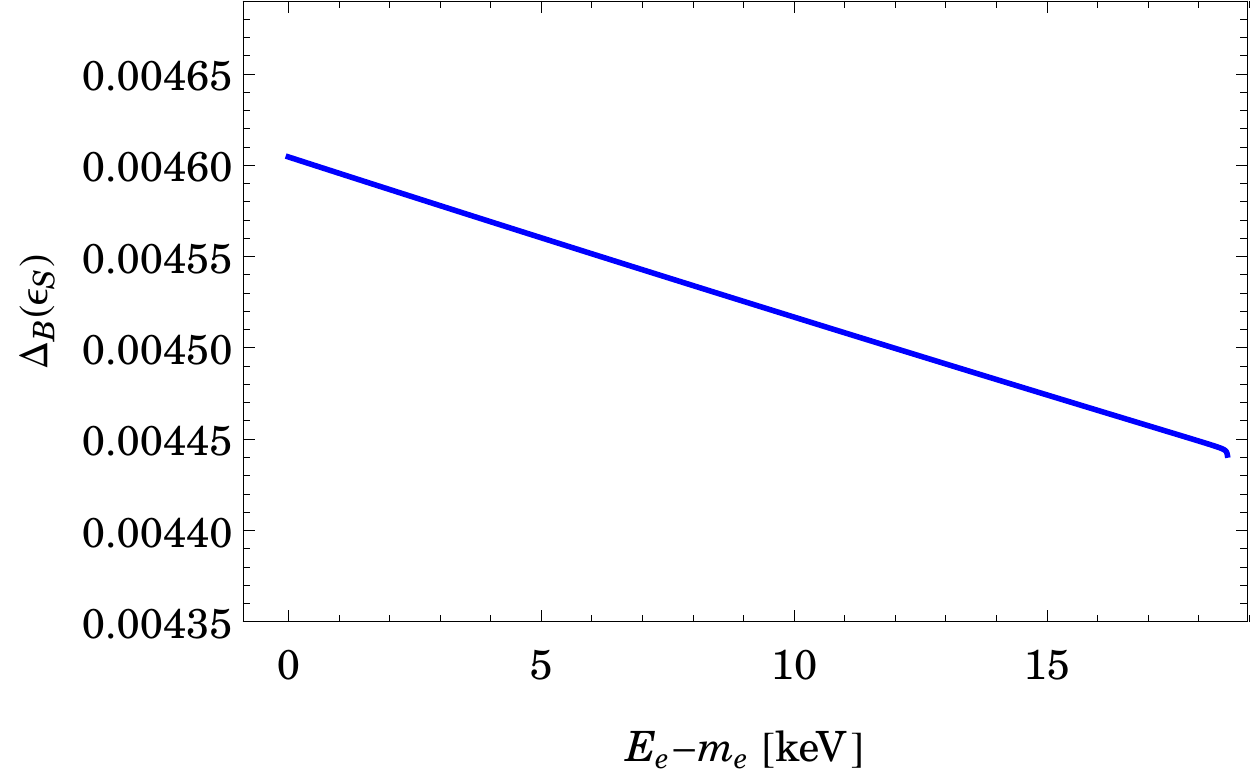}\hspace*{5mm}
\includegraphics[width=0.50\textwidth]{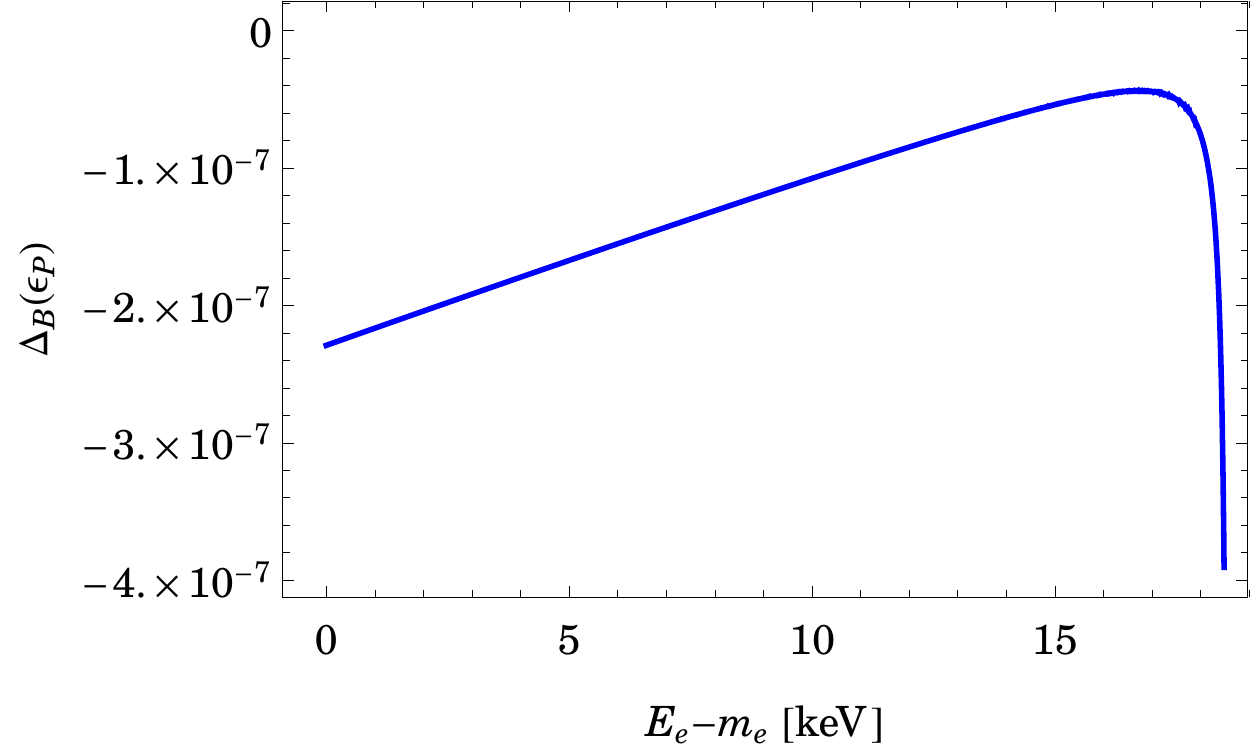}\\
\vspace*{3mm}
\includegraphics[width=0.50\textwidth]{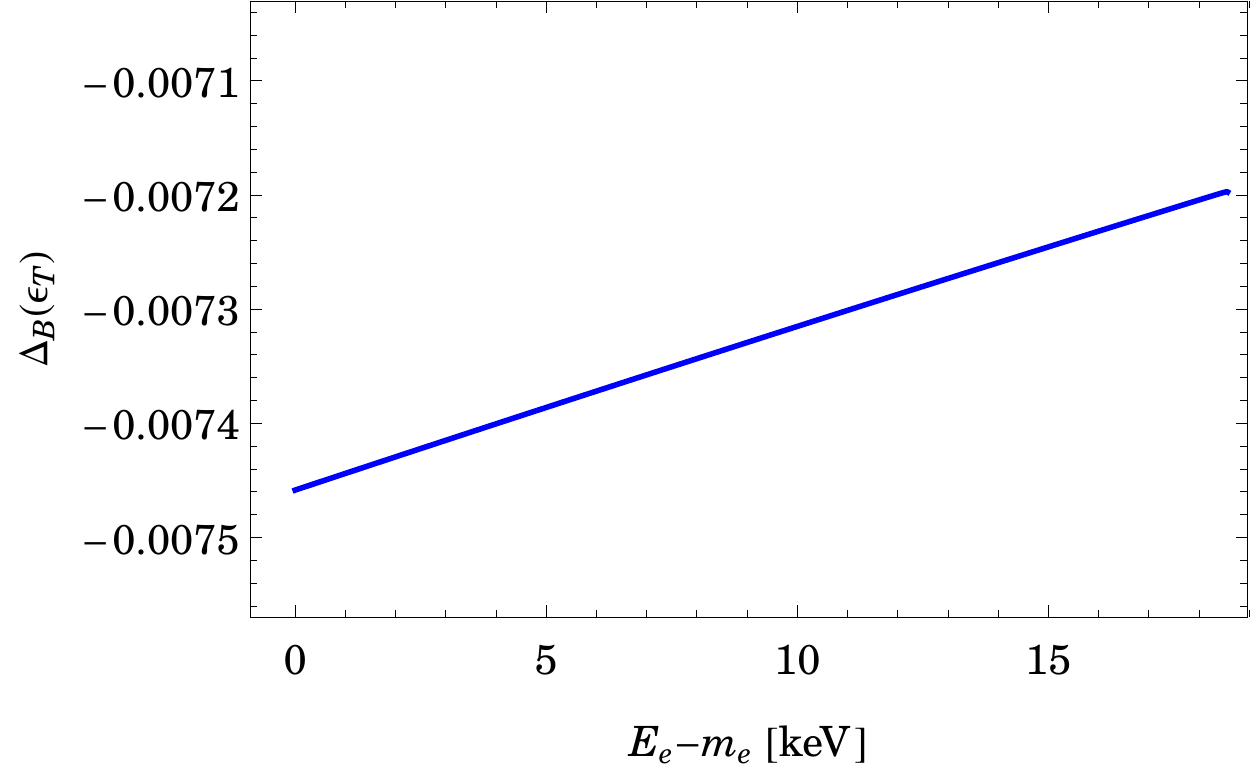}
\end{center}
\caption{Plots of the spectral distortions $\Delta_B(\stackrel{(\sim)}{\epsilon_j})$, see equation 
(\ref{eq:DB}), showing the
effect of new physics in the case of light active neutrinos with 
$m_\beta = 0.5$ eV. For smaller values of $m_\beta$ the plots look essentially the same.}
\label{comparisonB}
\end{figure}

\begin{figure}
\begin{center}
\includegraphics[width=0.50\textwidth]{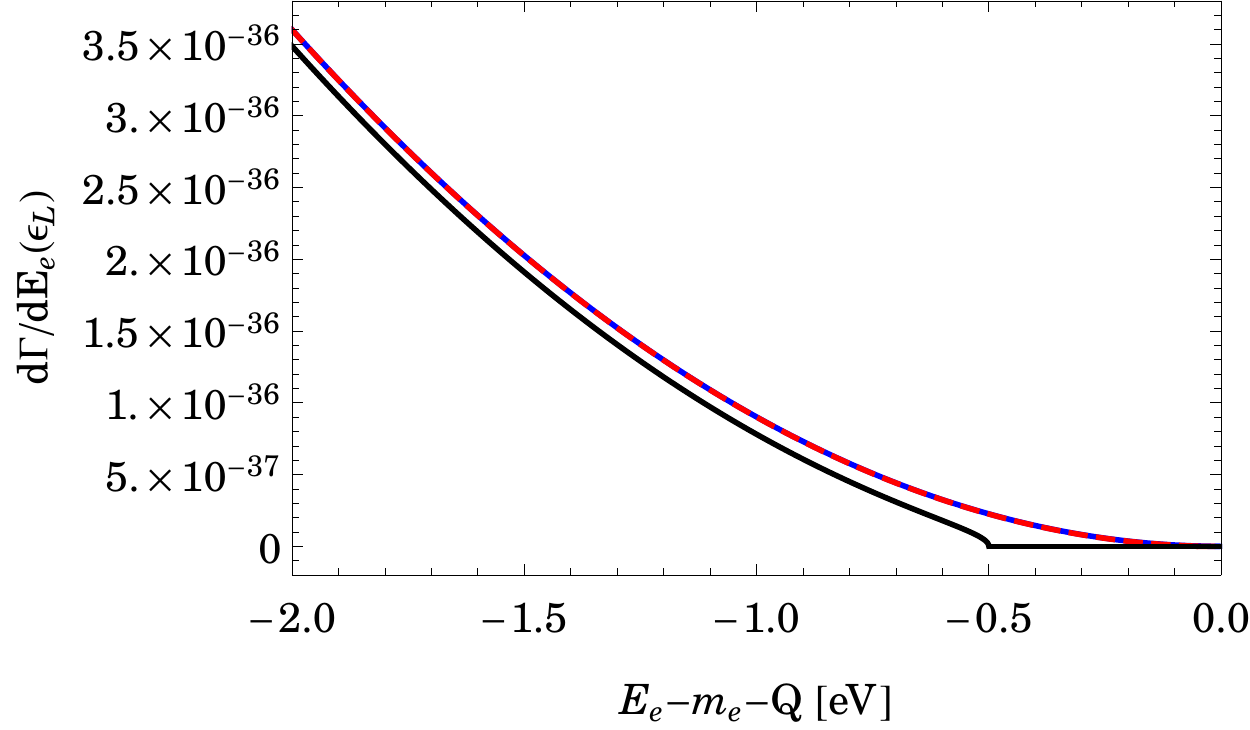}
\end{center}
\caption{The endpoints of the spectra AN (blue dashed), AI (red dashed) and B (black)
for $\epsilon_L=0.005$.} 
\label{endpointAB}
\end{figure}

\begin{figure}
\begin{center}
\includegraphics[width=0.50\textwidth]{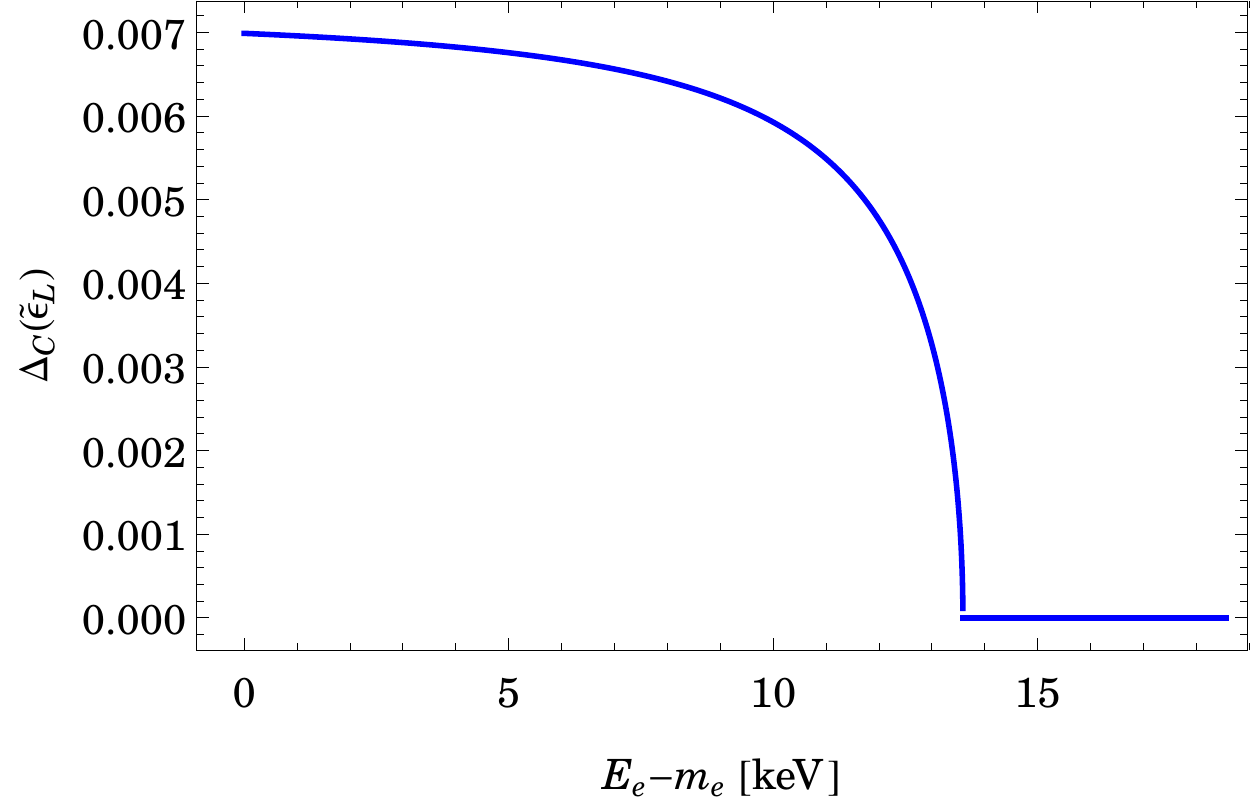}\hspace*{5mm}
\includegraphics[width=0.50\textwidth]{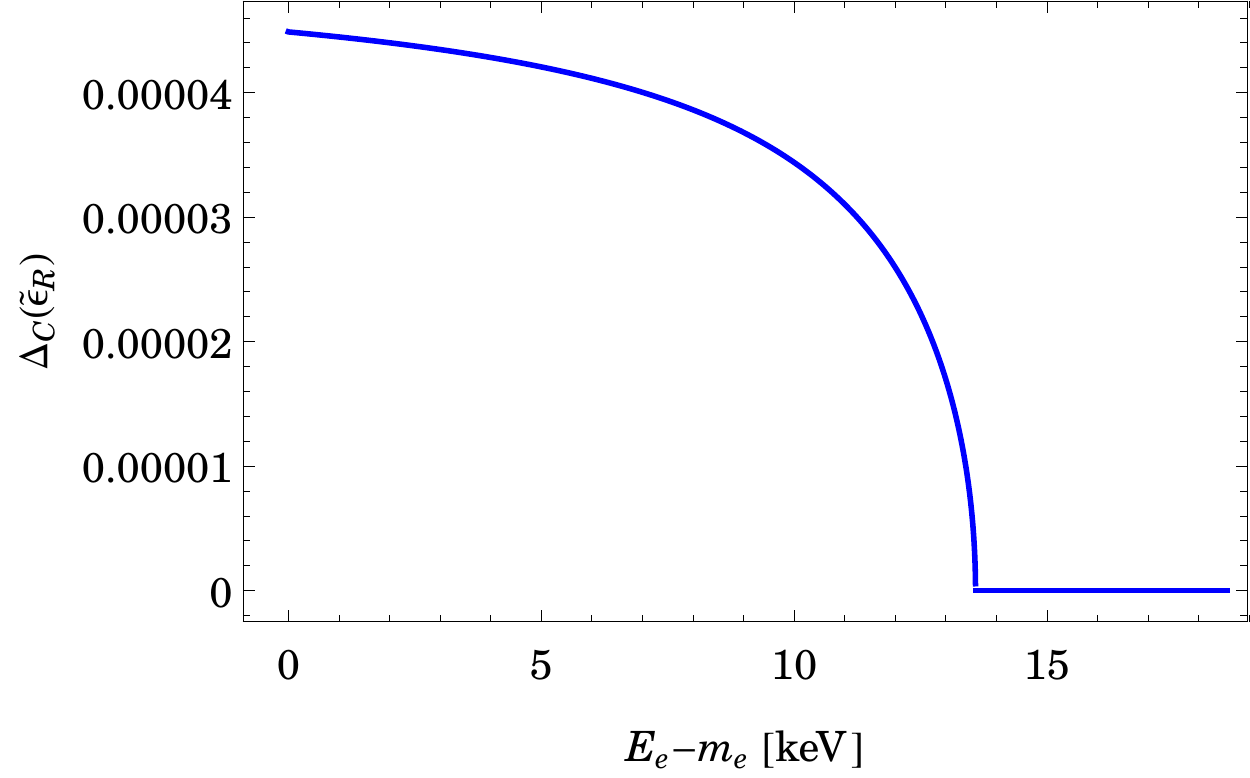}\\
\vspace*{3mm}
\includegraphics[width=0.50\textwidth]{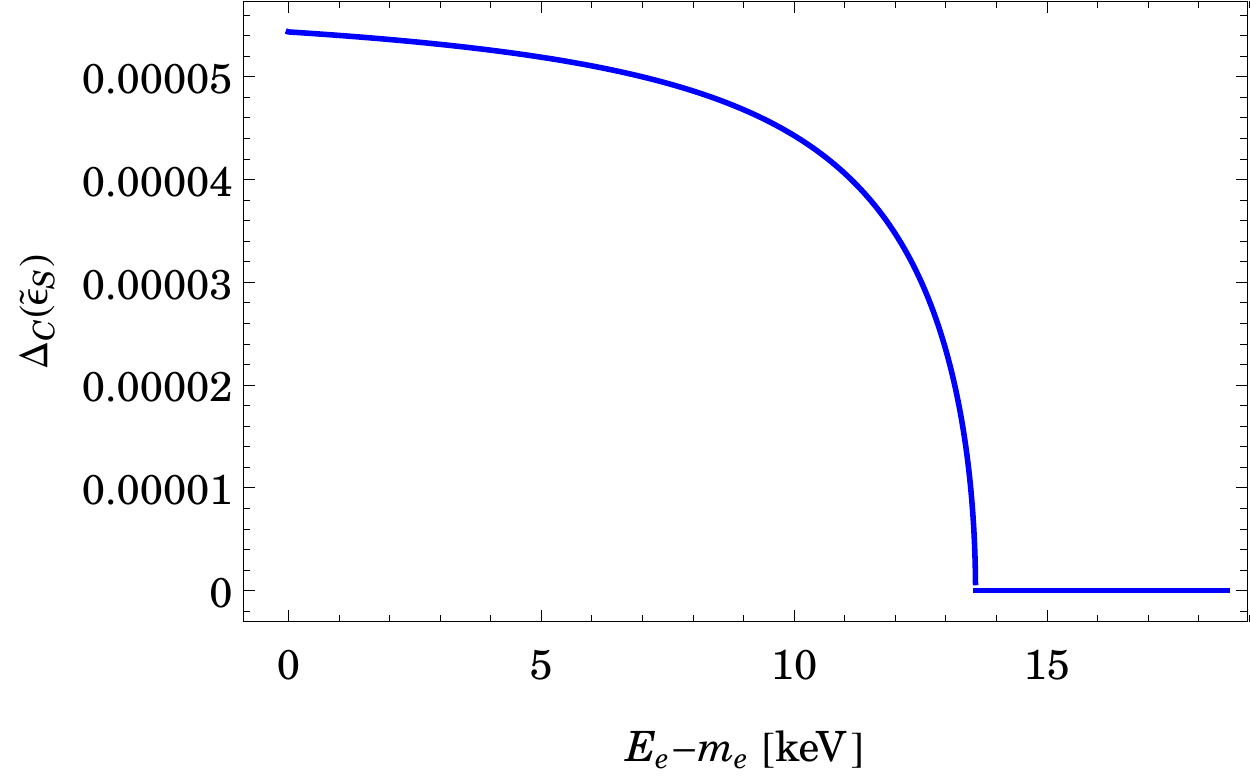}\hspace*{5mm}
\includegraphics[width=0.50\textwidth]{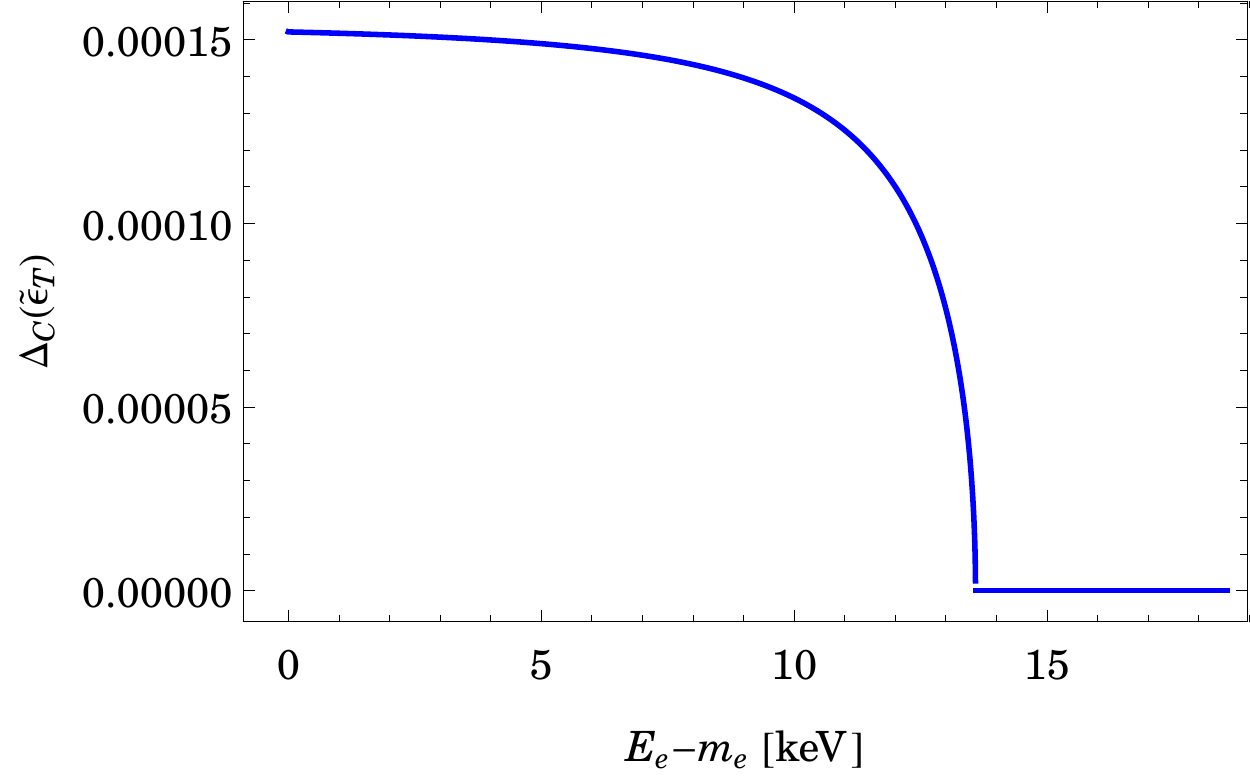}
\end{center}
\caption{Plots of the spectral distortions $\Delta_C(\widetilde{\epsilon}_j)$, see equation 
(\ref{eq:DC}), showing the
effect of new physics in the case of a sterile neutrino with 
$M_j = 5$ keV and mixing $S_{ei} = 10^{-3}$, $V_{ej}=1$.} 
\label{comparisonC}
\end{figure}

\begin{figure}
\begin{center}
\includegraphics[width=0.50\textwidth]{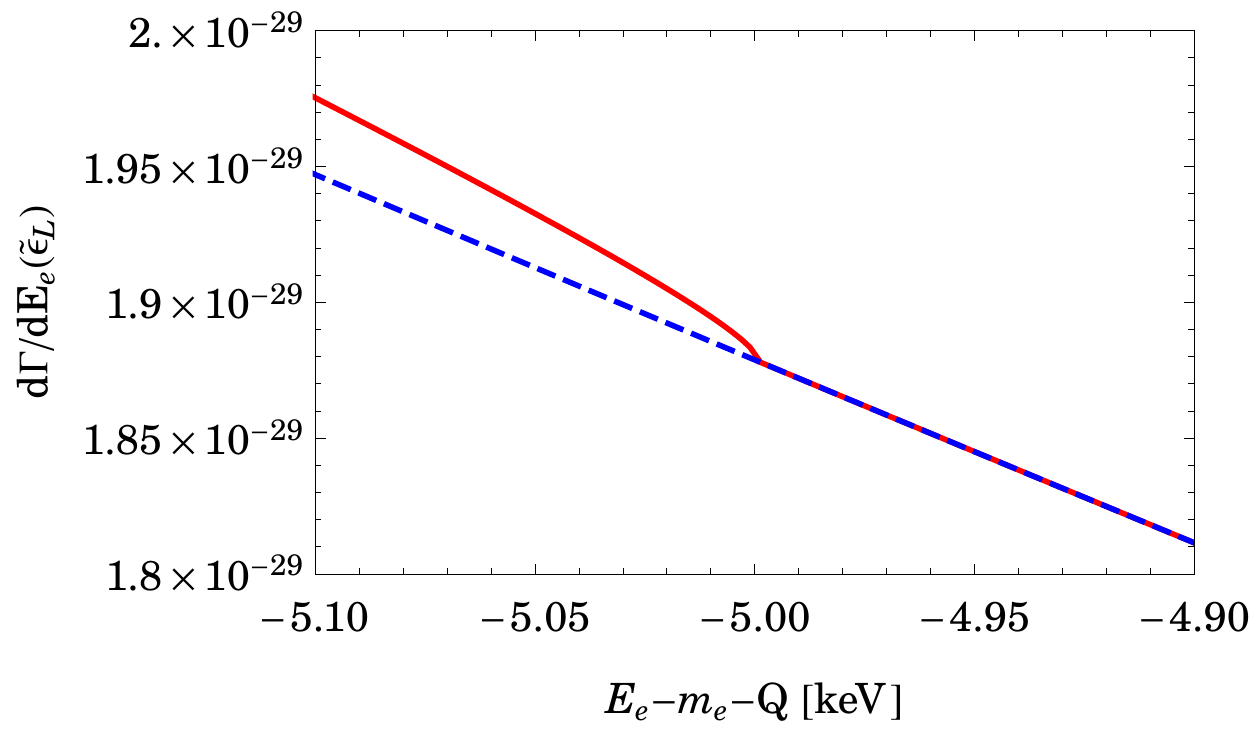}\hspace*{5mm}
\includegraphics[width=0.50\textwidth]{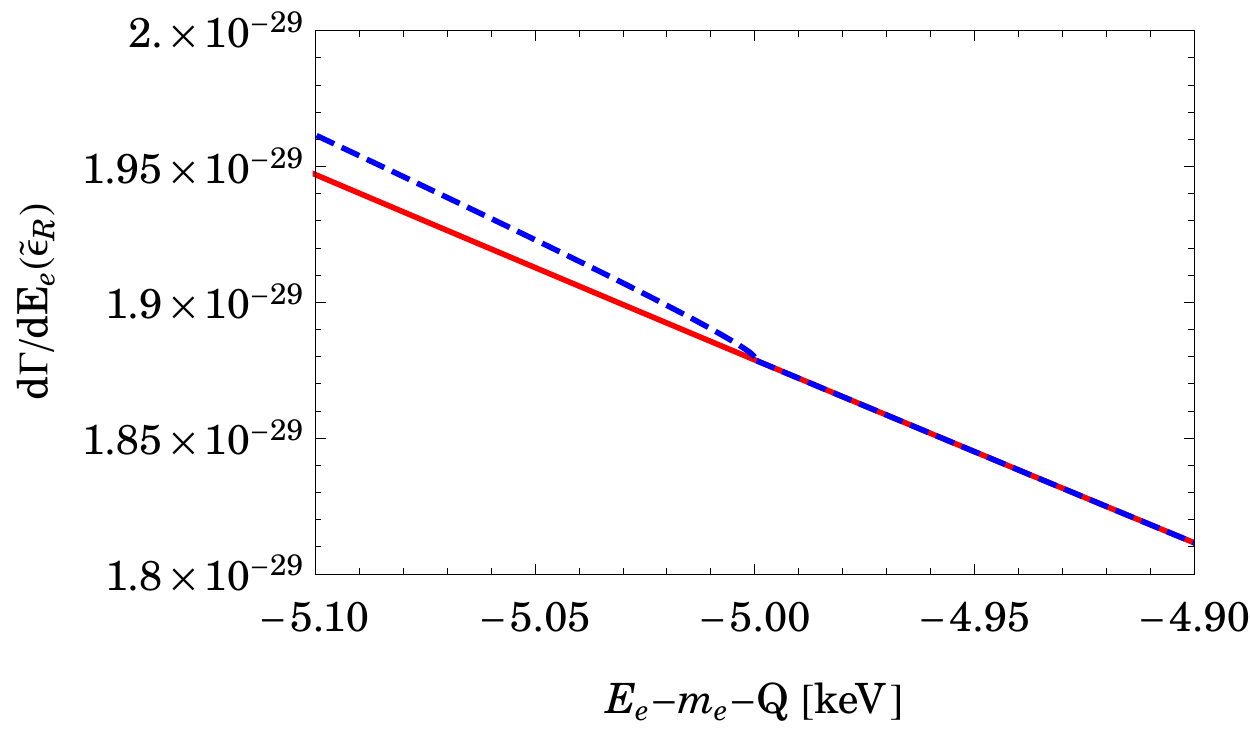}\\
\vspace*{3mm}
\includegraphics[width=0.50\textwidth]{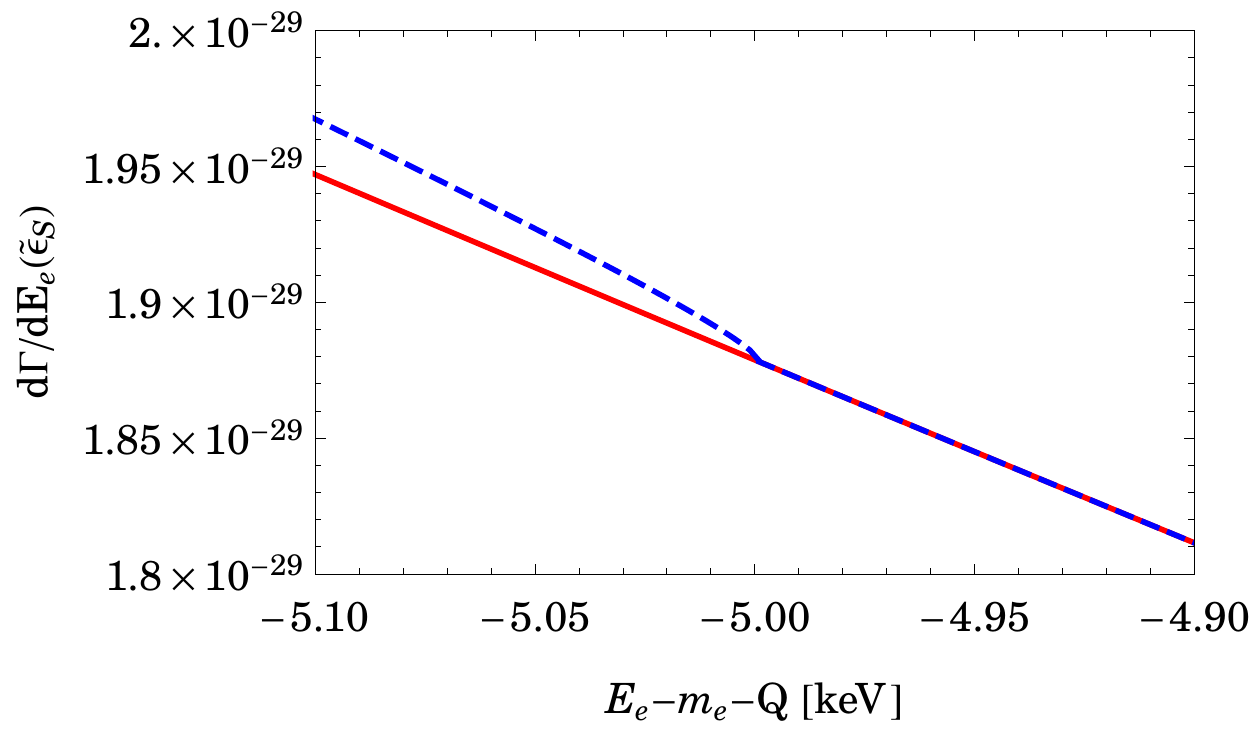}\hspace*{5mm}
\includegraphics[width=0.50\textwidth]{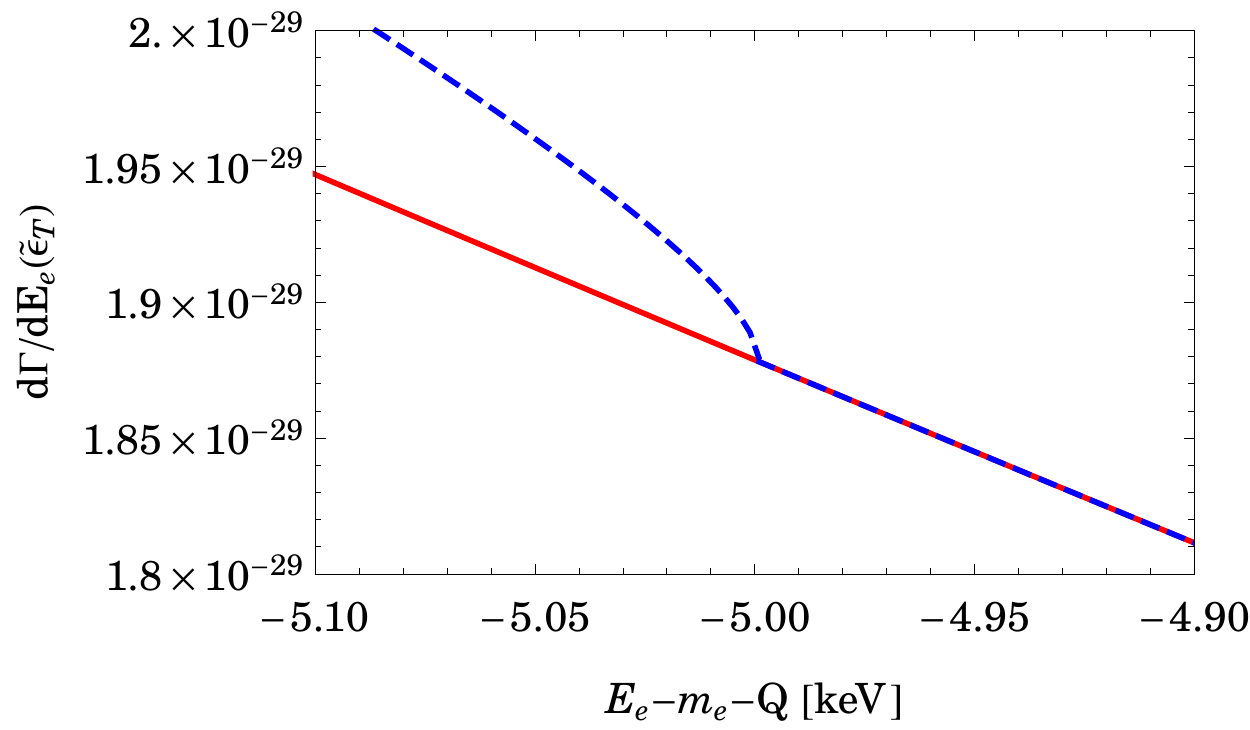}

\end{center}
\caption{The regions around the kink in the spectra of case C (heavy neutrino
with $M_j = 5$ keV and mixing $S_{ei} = 10^{-3}$, $V_{ej}=1$). In order to make the effects visible at this scale,
the contribution of the heavy neutrino to the total spectrum (blue dashed) has been multiplied by
a factor 10 (for $\widetilde{\epsilon}_L$)
and 1000 (for the other $\widetilde{\epsilon}$). The reference spectrum 2 (only light neutrinos) is shown in red.}
\label{endpointC}
\end{figure}

The resulting conclusions are as follows:\footnote{The new physics effects from 
$\epsilon_P$ and $\widetilde{\epsilon}_P$
are suppressed by $g_P \times q/M \ls 0.004$ and are therefore
expected to be almost negligible. Being negligible for heavy
neutrinos, in the case
of emission of light neutrinos the effect is probably also too small to be
observed. The effect
of pseudoscalar interactions is nevertheless shown in figure~\ref{comparisonB}.}
the plots AN (AI) \textit{vs.}\ 1N (1I) and B \textit{vs.}\ 2 are indistinguishable
both at the full scale ($E_e-m_e \in [0,\,Q]$) and also close to the endpoint
($E_e-m_e \in [Q-2\,\text{eV},\,Q]$). We therefore show only the plots for B \textit{vs.}\ 2
in figure~\ref{comparisonB}. 
To repeat, the quantity plotted is 
\begin{equation}\label{eq:DB}
\Delta_B(\stackrel{(\sim)}{\epsilon_\alpha}) 
 \equiv
\frac{\left(\frac{d\Gamma}{dE_e}\right)_{m_\beta = 0.5 \, \rm eV}^{\rm NP(\stackrel{(\sim)}{\epsilon_\alpha})}}{\left(\frac{d\Gamma}{dE_e}\right)_{m_\beta = 0.5 \, \rm eV}^{\rm no\,\, NP}} - 1 \,,
\end{equation}
\textit{i.e.}\ for quasi-degenerate neutrinos with $m_\beta = 0.5$ eV we show the relative 
ratio of the electron spectrum with and without new physics interactions governed by  
$\stackrel{(\sim)}{\epsilon_\alpha}$.
The endpoint plots for the different scenarios of new physics 
in cases A and B all look the same, \textit{i.e.}\ new physics has a negligible 
effect on the endpoint in the case of light neutrinos. For completeness, we 
show one of the endpoint plots (for $\epsilon_L$) in figure~\ref{endpointAB}. 
We can see however that interesting effects can be 
observed if the full 
spectrum is accessible. 
Also the case of heavy neutrinos as displayed in figure~\ref{comparisonC} shows 
interesting effects.
Again, we repeat that here the function 
\begin{equation}\label{eq:DC}
\Delta_C(\stackrel{(\sim)}{\epsilon_\alpha}) 
 \equiv
\frac{\left(\frac{d\Gamma}{dE_e}\right)_{M_j = 5 \, \rm keV}^{\rm NP(\stackrel{(\sim)}{\epsilon_\alpha})}}{\left(\frac{d\Gamma}{dE_e}\right)_{M_j = 5 \, \rm keV}^{\rm no\,\, NP}} - 1 
\end{equation}
is displayed, \textit{i.e.}\ for $m_\beta = 0.5$ eV and $M_j = 5$ keV with $S_{ej} = 10^{-3}$
we show the relative 
ratio of the electron spectrum with and without new physics interactions governed by  
$\stackrel{(\sim)}{\epsilon_\alpha}$. 
It is also worth to study the region of this function around the kink at $Q - M_j$, which is 
shown in figure~\ref{endpointC}.


\paragraph{Accessibility of the new-physics effects by a future KATRIN-like experiment:}
Without a dedicated experimental sensitivity study in terms of general 
spectral distortions, we have to satisfy ourselves with estimates on how 
current limits can be improved. 
In ref.~\cite{tritium-sensitivity} the effect of a heavy neutrino mass eigenstate in
the keV range on the spectrum is parameterized as
\begin{equation}
 \frac{d\Gamma}{dE_e} = \mathrm{cos}^2\theta \frac{d\Gamma}{dE_e}(m_\mathrm{light}) + \mathrm{sin}^2\theta \frac{d\Gamma}{dE_e}(m_\mathrm{heavy}),
\end{equation}
where the angle $\theta$ is a measure for active-sterile mixing. 
The potential sensitivities stated in~\cite{tritium-sensitivity} are, 
if the neutrino mass is not too small or too close to the endpoint, 
$\mathrm{sin}^2\theta<10^{-7}$ for the tritium source of KATRIN and
$\mathrm{sin}^2\theta<10^{-9}$ for a source with 100 times higher
activity. The final sensitivity that is achievable is still unclear at
the moment and requires further studies of theoretical corrections and
a full understanding of systematic effects in the experiment. 
We are aware that the modifications of the spectrum that are presented
here are not as obvious as ``simply'' a kink that is characteristic for a keV-scale
neutrino.  However, even in case of a keV neutrino the accompanying spectral
modification will be important to distinguish the signal from 
systematic effects~\cite{tritium-sensitivity}. 
In the present paper we will assume for illustration a somewhat optimistic sensitivity
of $10^{-7}$ on relative spectral distortions. 
Regarding the new physics interactions we can therefore estimate that 
if $\Delta(\stackrel{(\sim)}{\epsilon_j})$ exceeds 
$10^{-7}$ the effect will be observable. 
Another way to get a feeling for the testable effects is to consider the parameters 
$A,\ldots,D_2$ and assume that if 
$A/A_\mathrm{SM},\ldots, D_2/D_{2,\mathrm{SM}}$ exceeds $10^{-7}$ the effect is 
observable. See table~\ref{num2} for the numerical values of 
$A/A_\mathrm{SM},\ldots, D_2/D_{2,\mathrm{SM}}$, with numbers above $10^{-7}$ highlighted.

From figures \ref{comparisonB} and \ref{comparisonC} it becomes obvious 
that a modified KATRIN-like setup is sensitive to new physics 
represented by $\epsilon_L$, $\epsilon_R$, $\epsilon_S$, 
$\epsilon_T$ in case of 
left-handed light (even almost massless) 
neutrinos, while effects of $\epsilon_P$, though looking most
spectacular, are too small.  Those parameters appear linearly in the interference terms without 
suppression through the ratio of neutrino mass and electron energy, which is what happens 
for the $\widetilde \epsilon$. We can extract from the plots that 
the current limits on $\epsilon_{L}$, and $\epsilon_{R,S,T}$ could be improved by about 
six and five 
orders of magnitude, respectively. The limits can of course easily be rescaled
for less optimistic sensitivities than the one we use here $(10^{-7})$. 
Regarding the shape of the spectral distortion, 
the different figures offer means to distinguish the different $\epsilon_j$. 

For keV-scale right-handed neutrinos as displayed in figure \ref{comparisonC} it is the 
other way around, 
$\widetilde \epsilon_L$, $\widetilde \epsilon_R$, $\widetilde \epsilon_S$, 
$\widetilde \epsilon_T$ are of interest, while the interference terms 
involving $\epsilon$ are proportional to the ratio of 
neutrino 
mass over energy. 
This mass is for our example values in 
figures~\ref{comparisonB} and~\ref{comparisonC} a factor of $10^{4}$ larger, 
but the results for keV-neutrinos are suppressed by a factor of $10^{-6}$ from the mixing 
matrix elements $S$ and $T$, respectively. For our example values of the mixing, 
the bounds on 
$\widetilde \epsilon_L$, $\widetilde \epsilon_{R,S}$, and 
$\widetilde \epsilon_T$ could be improved by four, two and three orders of magnitude, though 
distinguishing them seems difficult. 

The main point to appreciate here is that the relative spectral
distortions of the electron energy spectrum can be on the permille
level for current limits on the exotic interactions. If understanding
of the theoretical uncertainties and control on systematical effects
beyond this level can be achieved, the current limits on the epsilon
parameters can be improved. 

%% file: conclusions.tex
\section{\label{sec:concl}Conclusions}

We have performed in this paper a study of the electron energy spectrum in nuclear $\beta$-decay, 
focussing on tritium because of the upcoming prospects of its investigation in 
KATRIN and other experiments. In particular, the full energy spectrum may be accessible, allowing 
the study of spectral distortions and additional heavy (sterile) neutrino mass states.  

First we have
carried out
a fully relativistic calculation of the 
spectrum, where we have demonstrated that in general the spectrum can
be parameterized by six 
functions which depend only on the involved particle masses and coupling constants. 
Those six functions are specified by the underlying interaction. Then we analyzed the 
spectrum in the Standard Model, discussing departures from
the non-relativistic results. 

Finally, using our relativistic calculation, 
we studied the potential spectral distortions in a general effective operator approach, 
taking all possible new charged current interactions into account, considering both 
light (sub-eV) and heavy (few keV) neutrinos. While the endpoint region does not display 
significant effects, the full spectrum can show
sizable distortions on the permille level, even for unobservably 
small neutrino masses. This allows in principle 
to improve the bounds on the effective operators and adds additional physics motivation to 
modifications of high activity neutrino mass experiments to study the full spectrum.

%% file: appendix.tex
\begin{appendix}

\section{Computation of $d\Gamma/dE_e$}\label{appA}
\setcounter{equation}{0}
\numberwithin{equation}{section}
Since in our special case $|\mathcal{M}|^2$ does not depend on the
direction of the emitted electron we can use
\begin{equation}
 \int \frac{d^3 p_e}{(2\pi)^3 \, 2E_e} =  \int \frac{4\pi\, d|\vec{p}_e| |\vec{p}_e|^2}{(2\pi)^3 \, 2E_e} = 
 \int \frac{4\pi\, dE_e\,E_e |\vec{p}_e|}{(2\pi)^3 \, 2E_e} = 
 \int \frac{dE_e\, \sqrt{E_e^2 - m_e^2}}{ 4\pi^2 }.
\end{equation}
The contribution to the electron spectrum is then given by
\begin{equation}
 \left( \frac{d\Gamma}{dE_e} \right)_{\overline{\nu_j}} = \frac{\sqrt{E_e^2-m_e^2}}{128\pi^4 m_\mathcal{A}}
 \int \frac{d^3 p_j d^3 p_\mathcal{B}}{E_j E_\mathcal{B}} 
 \left|\mathcal{M}(\mathcal{A} \rightarrow \mathcal{B} + e^- + \overline{\nu_j})\right|^2
  \delta^{(4)}(p_\mathcal{A}-p_\mathcal{B}-p_e-p_j).
\end{equation}
Since $|\mathcal{M}|^2$ does not depend on $\vec{p}_\mathcal{B}$ and $E_\mathcal{B}$
when given in the form of equation~(\ref{parameterization}), we can carry out the
$\vec{p}_\mathcal{B}$-integration and obtain
\begin{equation}\label{pj-int}
 \left( \frac{d\Gamma}{dE_e} \right)_{\overline{\nu_j}} = \frac{\sqrt{E_e^2-m_e^2}}{128\pi^4 m_\mathcal{A}}
 \int \frac{d^3 p_j}{E_j\, x} 
 \left|\mathcal{M}\right|^2
  \delta(m_\mathcal{A}-x-E_j-E_e),
\end{equation}
where
\begin{equation}
 x = \sqrt{|\vec{p}_e|^2 + |\vec{p}_j|^2 + 2|\vec{p}_e| |\vec{p}_j| \cos\vartheta + m_\mathcal{B}^2}.
\end{equation}
Here $\vartheta$ denotes the angle between $\vec{p}_e$ and $\vec{p}_j$.
Next we introduce polar coordinates for the $d^3p_j$-integration as follows:
\begin{equation}
 d^3 p_j =  d|\vec{p}_j| |\vec{p}_j|^2 d\varphi\, d\vartheta \, \sin\vartheta =
 dE_j E_j |\vec{p}_j| d\varphi\, d\vartheta \, \sin\vartheta.
\end{equation}
The integration over $\varphi$ gives a factor $2\pi$ and the integration over $\vartheta$
can be replaced by an integration over $x$~\cite{Griffiths}:
\begin{equation}
\frac{|\vec{p}_j| \sin\vartheta \, d\vartheta}{x} = -\frac{dx}{|\vec{p}_e|} = -\frac{dx}{\sqrt{E_e^2-m_e^2}}.
\end{equation}
We get
\begin{equation}
 \int \frac{d^3 p_j}{E_j\, x} 
 \left|\mathcal{M}\right|^2
  \delta(m_\mathcal{A}-x-E_j-E_e) = \frac{2\pi}{\sqrt{E_e^2-m_e^2}} \int dE_j \int_{x_-}^{x_+} dx
 \left|\mathcal{M}\right|^2
  \delta(m_\mathcal{A}-x-E_j-E_e),
\end{equation}
where
\begin{equation}
x_{\pm} = \sqrt{ (|\vec{p}_e| \pm |\vec{p}_j|)^2 + m_\mathcal{B}^2}.
\end{equation}
Since $|\mathcal{M}|^2$ does not depend on $x$, the $x$-integration reduces to
\begin{equation}
\int_{x_-}^{x_+} dx \, \delta(m_\mathcal{A}-x-E_j-E_e) =
\begin{cases}
1 & \text{if } m_\mathcal{A}-E_j-E_e \in (x_-,x_+) \\
0 & \text{else}
\end{cases}.
\end{equation}
The condition $m_\mathcal{A}-E_j-E_e \in (x_-,x_+)$ determines the
minimal and maximal electron energy\footnote{If $E_e \not\in [E_e^\text{min}, E_e^\text{max}]$,
$\left( \frac{d\Gamma}{dE_e} \right)_{\overline{\nu_j}}$ vanishes.}
\begin{equation}
E_e^\text{min} = m_e, \quad E_e^\text{max} = \frac{m_\mathcal{A}^2 + m_e^2 - (m_\mathcal{B} + m_j)^2}{2 m_\mathcal{A}}
\end{equation}
and the boundaries
\begin{equation}
E_{j\pm} = \frac{-(m_\mathcal{A}-E_e)(E_e m_\mathcal{A}-\alpha) \pm |\vec{p}_e|
\sqrt{(E_e m_\mathcal{A} -\alpha + m_j^2)^2 -m_\mathcal{B}^2 m_j^2}}{m_\mathcal{A}^2-2m_\mathcal{A}E_e + m_e^2}
\end{equation}
of the neutrino energy whose computation
is deferred to appendix~\ref{appB}.
The constant $\alpha$ is given by
\begin{equation}
\alpha = \frac{1}{2} \left( m_\mathcal{A}^2 - m_\mathcal{B}^2 + m_e^2 + m_j^2 \right).
\end{equation}
Thus, we have arrived at
\begin{equation}\label{x-integrated}
 \int \frac{d^3 p_j}{E_j\, x} 
 \left|\mathcal{M}\right|^2
  \delta(m_\mathcal{A}-x-E_j-E_e) = \frac{2\pi}{\sqrt{E_e^2-m_e^2}} \int_{E_{j-}}^{E_{j+}} dE_j
 \left|\mathcal{M}\right|^2.
\end{equation}
Inserting equation~(\ref{x-integrated}) into equation~(\ref{pj-int}) we find our final result
\begin{equation}
 \left( \frac{d\Gamma}{dE_e} \right)_{\overline{\nu_j}} = \frac{1}{64\pi^3 m_\mathcal{A}}
 \int_{E_{j-}}^{E_{j+}} dE_j
 \left|\mathcal{M}(\mathcal{A} \rightarrow \mathcal{B} + e^- + \overline{\nu_j})\right|^2.
\end{equation}

\section{Boundaries of the $E_j$-integration}\label{appB}
\setcounter{equation}{0}
The boundaries $E_{j-}$ and $E_{j+}$ of the $E_j$-integration
are determined by the three conditions
\begin{subequations}
\begin{align}
 & E_j>0,\label{c1}\\
 & m_\mathcal{A}-E_j-E_e > x_- = \sqrt{ (|\vec{p}_e| - |\vec{p}_j|)^2 + m_\mathcal{B}^2},\label{c2}\\
 & m_\mathcal{A}-E_j-E_e < x_+ = \sqrt{ (|\vec{p}_e| + |\vec{p}_j|)^2 + m_\mathcal{B}^2}.\label{c3}
\end{align}
\end{subequations}
In order to solve equations~(\ref{c2}) and~(\ref{c3}) for boundaries of $E_j$, we want to
square them. Doing so, we lose the condition $m_\mathcal{A}-E_j-E_e>0$ which follows from
inequality~(\ref{c2}). Thus, we have the two constraints
\begin{equation}\label{extra-constr}
 E_j>0, \quad E_j + E_e < m_\mathcal{A}
\end{equation}
and two further bounds on $E_j$ determined through
\begin{equation}\label{quad-ineq}
(m_\mathcal{A}-E_j-E_e)^2 > x_-^2 \quad\text{and}\quad (m_\mathcal{A}-E_j-E_e)^2 < x_+^2.
\end{equation}
The two inequalities~(\ref{quad-ineq}) may be written as a single inequality:
\begin{equation}\label{single-ineq}
\left(
\alpha - m_\mathcal{A} E_e - (m_\mathcal{A}-E_e) E_j
\right)^2 < |\vec{p}_e|^2 |\vec{p}_j|^2
\end{equation}
where we have defined
\begin{equation}
\alpha \equiv \frac{1}{2} \left( m_\mathcal{A}^2 - m_\mathcal{B}^2 + m_e^2 + m_j^2 \right).
\end{equation}
We may write equation~(\ref{single-ineq}) as
\begin{equation}\label{fineq}
f(E_e, E_j) < 0
\end{equation}
with
\begin{equation}
\begin{split}
 f(E_e, E_j) \equiv\enspace
& E_j^2 (m_\mathcal{A}^2 - 2 m_\mathcal{A} E_e + m_e^2 ) + \\
& E_j (-2 \alpha m_\mathcal{A} + 2 \alpha E_e + 2 m_\mathcal{A}^2 E_e - 2 m_\mathcal{A} E_e^2) + \\
& \alpha^2 + m_\mathcal{A}^2 E_e^2 -2 \alpha m_\mathcal{A} E_e + m_j^2 E_e^2 - m_e^2 m_j^2.
\end{split}
\end{equation}
The inequality~(\ref{fineq}) determines the boundaries of the kinematically allowed region
as $f(E_e, E_j) = 0$. Solving for $E_j$ gives two solutions
\begin{equation}
E_{j\pm} = \frac{-(m_\mathcal{A}-E_e)(E_e m_\mathcal{A}-\alpha) \pm |\vec{p}_e|
\sqrt{(E_e m_\mathcal{A} -\alpha + m_j^2)^2 -m_\mathcal{B}^2 m_j^2}}{m_\mathcal{A}^2-2m_\mathcal{A}E_e + m_e^2}.
\end{equation}
There are only these two solutions and they have no singularities for $E_e < \frac{m_\mathcal{A}^2+m_e^2}{2m_\mathcal{A}}$,
which is necessarily fulfilled for all beta decays.\footnote{Since $m_\mathcal{B}>m_\mathcal{A}/2$ we have
$E_e<m_\mathcal{A}/2<(m_\mathcal{A}^2+m_e^2)/2m_\mathcal{A}$.}
Moreover, $E_{j+}$ and $E_{j-}$ become equal at precisely the point where the term involving the square root vanishes.
Thus the two functions together describe the whole curve $f(E_e,E_j)=0$.

Furthermore, the two points where $E_{j+}=E_{j-}$ determine the maximal and minimal electron energy.
The minimum is given when $|\vec{p}_e| = 0$, \textit{i.e.} $E_e^\text{min} = m_e$. The maximum is reached
when the square root vanishes. This gives two solutions for $E_e^\text{max}$, namely
\begin{equation}
E_e^\text{max} = \frac{m_\mathcal{A}^2 + m_e^2 - (m_\mathcal{B} + m_j)^2}{2 m_\mathcal{A}}
\end{equation}
and the same solution with $(m_\mathcal{B} - m_j)^2$ instead of $(m_\mathcal{B} + m_j)^2$. However, this second solution
is unphysical, because it would correspond to a negative neutrino energy (which can be seen from
inserting the value for $E_e$ into the expression for $E_{j\pm}$).

It remains to check whether the bounds $E_{j\pm}$ are in agreement with the conditions of
equation~(\ref{extra-constr}).
The maximal and minimal neutrino energy (\textit{i.e.} the extrema of $E_{j+}$ and $E_{j-}$) can be
easily determined via $f(E_e, E_j)=0$ in the same way as we determined the maximal and minimal
electron energy. Since the whole problem is $e\leftrightarrow \overline{\nu}_j$-symmetric, any expression for the
neutrino can be obtained from the corresponding expression for the electron (and vice versa) via
the replacements
\begin{equation}
m_j \leftrightarrow m_e \quad\text{and}\quad E_j \leftrightarrow E_e.
\end{equation}
Thus the minimal and maximal neutrino energy are given by
\begin{equation}
 E_j^\text{min} = m_j, \quad E_j^\text{max} = \frac{m_\mathcal{A}^2 + m_j^2 - (m_\mathcal{B} + m_e)^2}{2 m_\mathcal{A}}.
\end{equation}
The condition $E_j>0$ is thus always fulfilled automatically, and also $E_j + E_e < m_\mathcal{A}$ is satisfied
due to
\begin{equation}
 E_j^\text{max} + E_e^\text{max} = \frac{1}{2 m_\mathcal{A}} ( 2m_\mathcal{A}^2 - 2 m_\mathcal{B}^2 - 2m_\mathcal{B} (m_e + m_j)) < m_\mathcal{A}.
\end{equation}
In total we have found that
\begin{equation}
\int_0^{\infty} dE_j \int_{x_-}^{x_+} dx
 \left|\mathcal{M}\right|^2
  \delta(m_\mathcal{A}-x-E_j-E_e) =
\begin{cases}
\displaystyle
\int_{E_{j-}}^{E_{j+}} dE_j  \left|\mathcal{M}\right|^2 & \text{for}\enspace E_e \in [E_e^\text{min}, E_e^\text{max}] \\
0 & \text{else}.
\end{cases}
\end{equation}

\end{appendix}

\begin{sidewaystable*}
\begin{scriptsize}
\begin{center}
\begin{tabular}{|c||c|c|c|c|c|c|c|c|c|c|}
\hline
 $\alpha$ & \multicolumn{2}{|c|}{$\epsilon_L$} & \multicolumn{2}{|c|}{$\epsilon_R$} & \multicolumn{2}{|c|}{$\epsilon_S$} & \multicolumn{2}{|c|}{$\epsilon_P$} & \multicolumn{2}{|c|}{$\epsilon_T$} \\
\hline
 & SM-NP & NP$^2$ & SM-NP & NP$^2$ & SM-NP & NP$^2$ & SM-NP & NP$^2$ & SM-NP & NP$^2$ \\
\hline
$A/A_\mathrm{SM}$ & $2$ & $1$ & $-8.72 \times 10^{0}$ & $1$ & $-3.11 \times 10^{-4}$ & $1.74 \times 10^{0}$ & $1.35 \times 10^{-1}$ & $1.92 \times 10^{1}$ & $1.62 \times 10^{-2}$ & $-2.79 \times 10^{1}$ \\
$B_1/B_{1,\mathrm{SM}}$ & $2$ & $1$ & $-8.72 \times 10^{0}$ & $9.92 \times 10^{-1}$ & $-3.11 \times 10^{-4}$ & $1.74 \times 10^{0}$ & $1.34 \times 10^{-1}$ & $3.84 \times 10^{1}$ & $1.62 \times 10^{-2}$ & $-2.79 \times 10^{1}$ \\
$B_2/B_{2,\mathrm{SM}}$ & $2$ & $1$ & $-8.79 \times 10^{0}$ & $1.01 \times 10^{0}$ & $3.13 \times 10^{-4}$ & $1.76 \times 10^{0}$ & $1.35 \times 10^{-1}$ & $3.88 \times 10^{1}$ & $6.78 \times 10^{-3}$ & $-2.81 \times 10^{1}$ \\
$C/C_\mathrm{SM}$ & $2$ & $1$ & $2$ & $1$ & $0$ & $-4.26 \times 10^{-2}$ & $0$ & $-4.99 \times 10^{3}$ & $1.86 \times 10^{-4}$ & $1.36 \times 10^{0}$ \\
$D_1/D_{1,\mathrm{SM}}$ & $2$ & $1$ & $-7.76 \times 10^{0}$ & $-6.82 \times 10^{0}$ & $-1.72 \times 10^{-4}$ & $1.58 \times 10^{-1}$ & $0$ & $1.84 \times 10^{4}$ & $-6.86 \times 10^{-4}$ & $5.04 \times 10^{0}$ \\
$D_2/D_{2,\mathrm{SM}}$ & $2$ & $1$ & $1.14 \times 10^{0}$ & $-1.47 \times 10^{-1}$ & $-2.52 \times 10^{-5}$ & $-2.31 \times 10^{-2}$ & $0$ & $-2.70 \times 10^{3}$ & $1.01 \times 10^{-4}$ & $-7.39 \times 10^{-1}$ \\
\hline
\end{tabular}
\bigskip
\\
\begin{tabular}{|c||c|c|c|c|c|c|c|c|c|c|}
\hline
 $\alpha$ & \multicolumn{2}{|c|}{$\widetilde{\epsilon}_L$} & \multicolumn{2}{|c|}{$\widetilde{\epsilon}_R$} & \multicolumn{2}{|c|}{$\widetilde{\epsilon}_S$} & \multicolumn{2}{|c|}{$\widetilde{\epsilon}_P$} & \multicolumn{2}{|c|}{$\widetilde{\epsilon}_T$} \\
\hline
$m_j=0.5\,\mathrm{eV}$ & SM-NP & NP$^2$ & SM-NP & NP$^2$ & SM-NP & NP$^2$ & SM-NP & NP$^2$ & SM-NP & NP$^2$ \\
\hline
$A/A_\mathrm{SM}$ & $-2.18 \times 10^{-9}$ & $1$ & $3.34 \times 10^{-9}$ & $1$ & $3.04 \times 10^{-10}$ & $1.74 \times 10^{0}$ & $-1.32 \times 10^{-7}$ & $1.92 \times 10^{1}$ & $6.59 \times 10^{-9}$ & $-2.79 \times 10^{1}$ \\
$B_1/B_{1,\mathrm{SM}}$ & $1.70 \times 10^{-12}$ & $9.92 \times 10^{-1}$ & $2.05 \times 10^{-12}$ & $1$ & $-3.04 \times 10^{-10}$ & $1.74 \times 10^{0}$ & $-1.31 \times 10^{-7}$ & $3.84 \times 10^{1}$ & $1.58 \times 10^{-8}$ & $-2.79 \times 10^{1}$ \\
$B_2/B_{2,\mathrm{SM}}$ & $1.72 \times 10^{-12}$ & $1.01 \times 10^{0}$ & $2.07 \times 10^{-12}$ & $1$ & $3.07 \times 10^{-10}$ & $1.76 \times 10^{0}$ & $-1.33 \times 10^{-7}$ & $3.88 \times 10^{1}$ & $6.64 \times 10^{-9}$ & $-2.81 \times 10^{1}$ \\
$C/C_\mathrm{SM}$ & $0$ & $1$ & $0$ & $1$ & $0$ & $-4.26 \times 10^{-2}$ & $0$ & $-4.99 \times 10^{3}$ & $1.82 \times 10^{-10}$ & $1.36 \times 10^{0}$ \\
$D_1/D_{1,\mathrm{SM}}$ & $0$ & $-6.82 \times 10^{0}$ & $0$ & $1$ & $-1.68 \times 10^{-10}$ & $1.58 \times 10^{-1}$ & $0$ & $1.84 \times 10^{4}$ & $-6.72 \times 10^{-10}$ & $5.04 \times 10^{0}$ \\
$D_2/D_{2,\mathrm{SM}}$ & $0$ & $-1.47 \times 10^{-1}$ & $0$ & $1$ & $-2.46 \times 10^{-11}$ & $-2.31 \times 10^{-2}$ & $0$ & $-2.70 \times 10^{3}$ & $9.84 \times 10^{-11}$ & $-7.39 \times 10^{-1}$ \\
\hline
\end{tabular}
\bigskip
\\
\begin{tabular}{|c||c|c|c|c|c|c|c|c|c|c|}
\hline
 $\alpha$ & \multicolumn{2}{|c|}{$\widetilde{\epsilon}_L$} & \multicolumn{2}{|c|}{$\widetilde{\epsilon}_R$} & \multicolumn{2}{|c|}{$\widetilde{\epsilon}_S$} & \multicolumn{2}{|c|}{$\widetilde{\epsilon}_P$} & \multicolumn{2}{|c|}{$\widetilde{\epsilon}_T$} \\
\hline
$M_j=5\,\mathrm{keV}$ & SM-NP & NP$^2$ & SM-NP & NP$^2$ & SM-NP & NP$^2$ & SM-NP & NP$^2$ & SM-NP & NP$^2$ \\
\hline
$A/A_\mathrm{SM}$ & $-2.18 \times 10^{-5}$ & $1$ & $3.34 \times 10^{-5}$ & $1$ & $3.04 \times 10^{-6}$ & $1.74 \times 10^{0}$ & $-1.32 \times 10^{-3}$ & $1.92 \times 10^{1}$ & $6.59 \times 10^{-5}$ & $-2.79 \times 10^{1}$ \\
$B_1/B_{1,\mathrm{SM}}$ & $1.70 \times 10^{-8}$ & $9.92 \times 10^{-1}$ & $2.05 \times 10^{-8}$ & $1$ & $-3.04 \times 10^{-6}$ & $1.74 \times 10^{0}$ & $-1.31 \times 10^{-3}$ & $3.84 \times 10^{1}$ & $1.58 \times 10^{-4}$ & $-2.79 \times 10^{1}$ \\
$B_2/B_{2,\mathrm{SM}}$ & $1.72 \times 10^{-8}$ & $1.01 \times 10^{0}$ & $2.07 \times 10^{-8}$ & $1$ & $3.07 \times 10^{-6}$ & $1.76 \times 10^{0}$ & $-1.33 \times 10^{-3}$ & $3.88 \times 10^{1}$ & $6.64 \times 10^{-5}$ & $-2.81 \times 10^{1}$ \\
$C/C_\mathrm{SM}$ & $0$ & $1$ & $0$ & $1$ & $0$ & $-4.26 \times 10^{-2}$ & $0$ & $-4.99 \times 10^{3}$ & $1.82 \times 10^{-6}$ & $1.36 \times 10^{0}$ \\
$D_1/D_{1,\mathrm{SM}}$ & $0$ & $-6.82 \times 10^{0}$ & $0$ & $1$ & $-1.68 \times 10^{-6}$ & $1.58 \times 10^{-1}$ & $0$ & $1.84 \times 10^{4}$ & $-6.72 \times 10^{-6}$ & $5.04 \times 10^{0}$ \\
$D_2/D_{2,\mathrm{SM}}$ & $0$ & $-1.47 \times 10^{-1}$ & $0$ & $1$ & $-2.46 \times 10^{-7}$ & $-2.31 \times 10^{-2}$ & $0$ & $-2.70 \times 10^{3}$ & $9.84 \times 10^{-7}$ & $-7.39 \times 10^{-1}$ \\
\hline
\end{tabular}
\caption{Numerical values for the coefficients $A$ to $D_2$ for $U_{ej}=V_{ej}=S_{ej}=T_{ej}=1$
and $\epsilon=\widetilde{\epsilon}=1$. We assumed only one massive neutrino.
The upper table (for $\epsilon$) is valid for both $m_j=0.5\,\mathrm{eV}$ and $M_j=5\,\mathrm{keV}$.
The lower two tables (for $\widetilde{\epsilon}$) are for $m_j=0.5\,\mathrm{eV}$ (middle table)
and $M_j=5\,\mathrm{keV}$ (lowest table).
SM-NP and NP$^2$ stand for $2\,\mathrm{Re}\,S_{\mathrm{SM},\alpha}$ and $S_{\alpha\alpha}$, respectively.
The values for $A,\ldots,D_2$ can be obtained by multiplying the values given in the table with the values
for the SM contribution and the corresponding suppression factor from table~\ref{suppression}.
The SM-NP interference contributions from $\widetilde{\epsilon}$
vanish for $m_j\rightarrow 0$ and are consequently suppressed by the smallness of the neutrino mass.
Therefore the SM-NP terms of the two lower tables are related by a factor of $0.5\,\mathrm{eV}/5\,\mathrm{keV}=10^{-4}$.
The NP$^2$ terms of the two lower tables are equal.}\label{num1}
\end{center}
\end{scriptsize}
\end{sidewaystable*}

\begin{sidewaystable*}
\begin{scriptsize}
\begin{center}
\begin{tabular}{|c||c|c|c|c|c|c|c|c|c|c|}
\hline
$\alpha$ & \multicolumn{2}{|c|}{$\epsilon_L$} & \multicolumn{2}{|c|}{$\epsilon_R$} & \multicolumn{2}{|c|}{$\epsilon_S$} & \multicolumn{2}{|c|}{$\epsilon_P$} & \multicolumn{2}{|c|}{$\epsilon_T$} \\
\hline
$m_j=0.5\,\mathrm{eV}$  & SM-NP & NP$^2$ & SM-NP & NP$^2$ & SM-NP & NP$^2$ & SM-NP & NP$^2$ & SM-NP & NP$^2$ \\
\hline
$A/A_\mathrm{SM}$ & \col{$1.00 \times 10^{-2}$} & \col{$2.50 \times 10^{-5}$} & \col{$-6.19 \times 10^{-3}$} & \col{$5.04 \times 10^{-7}$} & \col{$-4.04 \times 10^{-6}$} & \col{$2.95 \times 10^{-4}$} & \col{$6.05 \times 10^{-5}$} & \col{$3.90 \times 10^{-6}$} & \col{$2.26 \times 10^{-5}$} & \col{$-5.47 \times 10^{-5}$} \\
$B_1/B_{1,\mathrm{SM}}$ & \col{$1.00 \times 10^{-2}$} & \col{$2.50 \times 10^{-5}$} & \col{$-6.19 \times 10^{-3}$} & \col{$5.00 \times 10^{-7}$} & \col{$-4.04 \times 10^{-6}$} & \col{$2.94 \times 10^{-4}$} & \col{$6.05 \times 10^{-5}$} & \col{$7.78 \times 10^{-6}$} & \col{$2.26 \times 10^{-5}$} & \col{$-5.46 \times 10^{-5}$} \\
$B_2/B_{2,\mathrm{SM}}$ & \col{$1.00 \times 10^{-2}$} & \col{$2.50 \times 10^{-5}$} & \col{$-6.24 \times 10^{-3}$} & \col{$5.08 \times 10^{-7}$} & \col{$4.07 \times 10^{-6}$} & \col{$2.97 \times 10^{-4}$} & \col{$6.10 \times 10^{-5}$} & \col{$7.85 \times 10^{-6}$} & \col{$9.50 \times 10^{-6}$} & \col{$-5.51 \times 10^{-5}$} \\
$C/C_\mathrm{SM}$ & \col{$1.00 \times 10^{-2}$} & \col{$2.50 \times 10^{-5}$} & \col{$1.42 \times 10^{-3}$} & \col{$5.04 \times 10^{-7}$} & $0$ & \col{$-7.20 \times 10^{-6}$} & $0$ & \col{$-1.01 \times 10^{-3}$} & \col{$2.60 \times 10^{-7}$} & \col{$2.67 \times 10^{-6}$} \\
$D_1/D_{1,\mathrm{SM}}$ & \col{$1.00 \times 10^{-2}$} & \col{$2.50 \times 10^{-5}$} & \col{$-5.51 \times 10^{-3}$} & \col{$-3.44 \times 10^{-6}$} & \col{$-2.23 \times 10^{-6}$} & \col{$2.66 \times 10^{-5}$} & $0$ & \col{$3.74 \times 10^{-3}$} & \col{$-9.61 \times 10^{-7}$} & \col{$9.88 \times 10^{-6}$} \\
$D_2/D_{2,\mathrm{SM}}$ & \col{$1.00 \times 10^{-2}$} & \col{$2.50 \times 10^{-5}$} & \col{$8.08 \times 10^{-4}$} & $-7.39 \times 10^{-8}$ & \col{$-3.27 \times 10^{-7}$} & \col{$-3.90 \times 10^{-6}$} & $0$ & \col{$-5.47 \times 10^{-4}$} & \col{$1.41 \times 10^{-7}$} & \col{$-1.45 \times 10^{-6}$} \\
\hline
\end{tabular}
\bigskip
\\
\begin{tabular}{|c||c|c|c|c|c|c|c|c|c|c|}
\hline
 $\alpha$ & \multicolumn{2}{|c|}{$\widetilde{\epsilon}_L$} & \multicolumn{2}{|c|}{$\widetilde{\epsilon}_R$} & \multicolumn{2}{|c|}{$\widetilde{\epsilon}_S$} & \multicolumn{2}{|c|}{$\widetilde{\epsilon}_P$} & \multicolumn{2}{|c|}{$\widetilde{\epsilon}_T$} \\
\hline
$m_j=0.5\,\mathrm{eV}$ & SM-NP & NP$^2$ & SM-NP & NP$^2$ & SM-NP & NP$^2$ & SM-NP & NP$^2$ & SM-NP & NP$^2$ \\
\hline
$A/A_\mathrm{SM}$ & $-1.86 \times 10^{-13}$ & $7.22 \times 10^{-9}$ & $2.37 \times 10^{-14}$ & $5.04 \times 10^{-11}$ & $5.48 \times 10^{-15}$ & $5.65 \times 10^{-10}$ & $-3.69 \times 10^{-14}$ & $1.51 \times 10^{-12}$ & $2.77 \times 10^{-14}$ & $-4.92 \times 10^{-10}$ \\
$B_1/B_{1,\mathrm{SM}}$ & $1.45 \times 10^{-16}$ & $7.17 \times 10^{-9}$ & $1.46 \times 10^{-17}$ & $5.04 \times 10^{-11}$ & $-5.47 \times 10^{-15}$ & $5.64 \times 10^{-10}$ & $-3.68 \times 10^{-14}$ & $3.01 \times 10^{-12}$ & $6.64 \times 10^{-14}$ & $-4.91 \times 10^{-10}$ \\
$B_2/B_{2,\mathrm{SM}}$ & $1.46 \times 10^{-16}$ & $7.28 \times 10^{-9}$ & $1.47 \times 10^{-17}$ & $5.04 \times 10^{-11}$ & $5.52 \times 10^{-15}$ & $5.69 \times 10^{-10}$ & $-3.71 \times 10^{-14}$ & $3.04 \times 10^{-12}$ & $2.79 \times 10^{-14}$ & $-4.95 \times 10^{-10}$ \\
$C/C_\mathrm{SM}$ & $0$ & $7.23 \times 10^{-9}$ & $0$ & $5.04 \times 10^{-11}$ & $0$ & $-1.38 \times 10^{-11}$ & $0$ & $-3.91 \times 10^{-10}$ & $7.63 \times 10^{-16}$ & $2.40 \times 10^{-11}$ \\
$D_1/D_{1,\mathrm{SM}}$ & $0$ & $-4.93 \times 10^{-8}$ & $0$ & $5.04 \times 10^{-11}$ & $-3.02 \times 10^{-15}$ & $5.11 \times 10^{-11}$ & $0$ & $1.45 \times 10^{-9}$ & $-2.82 \times 10^{-15}$ & $8.89 \times 10^{-11}$ \\
$D_2/D_{2,\mathrm{SM}}$ & $0$ & $-1.06 \times 10^{-9}$ & $0$ & $5.04 \times 10^{-11}$ & $-4.43 \times 10^{-16}$ & $-7.48 \times 10^{-12}$ & $0$ & $-2.12 \times 10^{-10}$ & $4.13 \times 10^{-16}$ & $-1.30 \times 10^{-11}$ \\
\hline
\end{tabular}
\bigskip
\\
\begin{tabular}{|c||c|c|c|c|c|c|c|c|c|c|}
\hline
$\alpha$ & \multicolumn{2}{|c|}{$\epsilon_L$} & \multicolumn{2}{|c|}{$\epsilon_R$} & \multicolumn{2}{|c|}{$\epsilon_S$} & \multicolumn{2}{|c|}{$\epsilon_P$} & \multicolumn{2}{|c|}{$\epsilon_T$} \\
\hline
$M_j=5\,\mathrm{keV}$ & SM-NP & NP$^2$ & SM-NP & NP$^2$ & SM-NP & NP$^2$ & SM-NP & NP$^2$ & SM-NP & NP$^2$ \\
\hline
$A/A_\mathrm{SM}$ & $1.00 \times 10^{-8}$ & $2.50 \times 10^{-11}$ & $-6.19 \times 10^{-9}$ & $5.04 \times 10^{-13}$ & $-4.04 \times 10^{-12}$ & $2.95 \times 10^{-10}$ & $6.05 \times 10^{-11}$ & $3.90 \times 10^{-12}$ & $2.26 \times 10^{-11}$ & $-5.47 \times 10^{-11}$ \\
$B_1/B_{1,\mathrm{SM}}$ & $1.00 \times 10^{-8}$ & $2.50 \times 10^{-11}$ & $-6.19 \times 10^{-9}$ & $5.00 \times 10^{-13}$ & $-4.04 \times 10^{-12}$ & $2.94 \times 10^{-10}$ & $6.05 \times 10^{-11}$ & $7.78 \times 10^{-12}$ & $2.26 \times 10^{-11}$ & $-5.46 \times 10^{-11}$ \\
$B_2/B_{2,\mathrm{SM}}$ & $1.00 \times 10^{-8}$ & $2.50 \times 10^{-11}$ & $-6.24 \times 10^{-9}$ & $5.08 \times 10^{-13}$ & $4.07 \times 10^{-12}$ & $2.97 \times 10^{-10}$ & $6.10 \times 10^{-11}$ & $7.85 \times 10^{-12}$ & $9.50 \times 10^{-12}$ & $-5.51 \times 10^{-11}$ \\
$C/C_\mathrm{SM}$ & $1.00 \times 10^{-8}$ & $2.50 \times 10^{-11}$ & $1.42 \times 10^{-9}$ & $5.04 \times 10^{-13}$ & $0$ & $-7.20 \times 10^{-12}$ & $0$ & $-1.01 \times 10^{-9}$ & $2.60 \times 10^{-13}$ & $2.67 \times 10^{-12}$ \\
$D_1/D_{1,\mathrm{SM}}$ & $1.00 \times 10^{-8}$ & $2.50 \times 10^{-11}$ & $-5.51 \times 10^{-9}$ & $-3.44 \times 10^{-12}$ & $-2.23 \times 10^{-12}$ & $2.66 \times 10^{-11}$ & $0$ & $3.74 \times 10^{-9}$ & $-9.61 \times 10^{-13}$ & $9.88 \times 10^{-12}$ \\
$D_2/D_{2,\mathrm{SM}}$ & $1.00 \times 10^{-8}$ & $2.50 \times 10^{-11}$ & $8.08 \times 10^{-10}$ & $-7.39 \times 10^{-14}$ & $-3.27 \times 10^{-13}$ & $-3.90 \times 10^{-12}$ & $0$ & $-5.47 \times 10^{-10}$ & $1.41 \times 10^{-13}$ & $-1.45 \times 10^{-12}$ \\
\hline
\end{tabular}
\bigskip
\\
\begin{tabular}{|c||c|c|c|c|c|c|c|c|c|c|}
\hline
 $\alpha$ & \multicolumn{2}{|c|}{$\widetilde{\epsilon}_L$} & \multicolumn{2}{|c|}{$\widetilde{\epsilon}_R$} & \multicolumn{2}{|c|}{$\widetilde{\epsilon}_S$} & \multicolumn{2}{|c|}{$\widetilde{\epsilon}_P$} & \multicolumn{2}{|c|}{$\widetilde{\epsilon}_T$} \\
\hline
$M_j=5\,\mathrm{keV}$  & SM-NP & NP$^2$ & SM-NP & NP$^2$ & SM-NP & NP$^2$ & SM-NP & NP$^2$ & SM-NP & NP$^2$ \\
\hline
$A/A_\mathrm{SM}$ & $-1.86 \times 10^{-9}$ & \col{$7.23 \times 10^{-3}$} & $2.37 \times 10^{-10}$ & \col{$5.04 \times 10^{-5}$} & $5.48 \times 10^{-11}$ & \col{$5.65 \times 10^{-4}$} & $-3.69 \times 10^{-10}$ & \col{$1.51 \times 10^{-6}$} & $2.77 \times 10^{-10}$ & \col{$-4.92 \times 10^{-4}$} \\
$B_1/B_{1,\mathrm{SM}}$ & $1.45 \times 10^{-12}$ & \col{$7.17 \times 10^{-3}$} & $1.46 \times 10^{-13}$ & \col{$5.04 \times 10^{-5}$} & $-5.47 \times 10^{-11}$ & \col{$5.64 \times 10^{-4}$} & $-3.68 \times 10^{-10}$ & \col{$3.01 \times 10^{-6}$} & $6.64 \times 10^{-10}$ & \col{$-4.91 \times 10^{-4}$} \\
$B_2/B_{2,\mathrm{SM}}$ & $1.46 \times 10^{-12}$ & \col{$7.28 \times 10^{-3}$} & $1.47 \times 10^{-13}$ & \col{$5.04 \times 10^{-5}$} & $5.52 \times 10^{-11}$ & \col{$5.69 \times 10^{-4}$} & $-3.71 \times 10^{-10}$ & \col{$3.04 \times 10^{-6}$} & $2.79 \times 10^{-10}$ & \col{$-4.95 \times 10^{-4}$} \\
$C/C_\mathrm{SM}$ & $0$ & \col{$7.23 \times 10^{-3}$} & $0$ & \col{$5.04 \times 10^{-5}$} & $0$ & \col{$-1.38 \times 10^{-5}$} & $0$ & \col{$-3.91 \times 10^{-4}$} & $7.63 \times 10^{-12}$ & \col{$2.40 \times 10^{-5}$} \\
$D_1/D_{1,\mathrm{SM}}$ & $0$ & \col{$-4.93 \times 10^{-2}$} & $0$ & \col{$5.04 \times 10^{-5}$} & $-3.02 \times 10^{-11}$ & \col{$5.11 \times 10^{-5}$} & $0$ & \col{$1.45 \times 10^{-3}$} & $-2.82 \times 10^{-11}$ & \col{$8.89 \times 10^{-5}$} \\
$D_2/D_{2,\mathrm{SM}}$ & $0$ & \col{$-1.06 \times 10^{-3}$} & $0$ & \col{$5.04 \times 10^{-5}$} & $-4.43 \times 10^{-12}$ & \col{$-7.48 \times 10^{-6}$} & $0$ & \col{$-2.12 \times 10^{-4}$} & $4.13 \times 10^{-12}$ & \col{$-1.30 \times 10^{-5}$} \\
\hline
\end{tabular}
\caption{The numerical estimates for the size of new physics effects in tritium beta decay
compared to the Standard Model. The first and third table are equal up to the factor $|S_{ej}/U_{ej}|^2=|T_{ej}/V_{ej}|^2=10^{-6}$.
The SM-NP contributions in the second and fourth table show the suppression
by the neutrino mass $0.5\,\text{eV} / 5\,\text{keV}=10^{-4}$, while the NP$^2$ contributions
show the $|S_{ej}/U_{ej}|^2=|T_{ej}/V_{ej}|^2=10^{-6}$ suppression. All numbers with absolute value larger than
$10^{-7}$ are highlighted.}\label{num2}
\end{center}
\end{scriptsize}
\end{sidewaystable*}

\newpage